\documentclass[journal=nalefd,manuscript=letter]{achemso}
\usepackage{amsmath}
\usepackage{amssymb}
\usepackage{bm}
\usepackage{graphicx}
\usepackage{times}
\usepackage{color}
\usepackage{braket}
\usepackage{ulem}
\let\vec=\mathbf

\title{Bound states in the continuum \\ in photonic structures}

\author{Kirill Koshelev}
\affiliation{School of Physics and Engineering, ITMO University, 197101 St.~Petersburg, Russian Federation}
\alsoaffiliation{Nonlinear Physics Center, Research School of Physics, Australian National University, Canberra ACT 2601, Australia}
\author{Zarina Sadrieva}
\affiliation{School of Physics and Engineering, ITMO University, 197101 St.~Petersburg, Russian Federation}
\author{Alexey Shcherbakov}
\affiliation{School of Physics and Engineering, ITMO University, 197101 St.~Petersburg, Russian Federation}
\author{Yuri Kivshar}
\affiliation{School of Physics and Engineering, ITMO University, 197101 St.~Petersburg, Russian Federation}
\alsoaffiliation{Nonlinear Physics Center, Research School of Physics, Australian National University, Canberra ACT 2601, Australia}
\author{Andrey Bogdanov}
\affiliation{School of Physics and Engineering, ITMO University, 197101 St.~Petersburg, Russian Federation}
\email{a.bogdanov@metalab.ifmo.ru}
\date{\today}

\begin{document}

\begin{abstract}
Bound states in the continuum provide a remarkable example of how a simple problem solved about a century ago in quantum mechanics can drive the research on a whole spectrum of resonant phenomena in wave physics. Due to their huge radiative lifetime, bound states in the continuum have found multiple applications in various areas of physics devoted to wave processes, including hydrodynamics, atomic physics, and acoustics. In this review paper, we present a comprehensive description of bound states in the continuum and related effects, focusing mainly on photonic dielectric structures. We review the history of this area, basic physical mechanisms in the formation of bound states in the continuum, and specific examples of structures supporting such states. We also discuss their possible applications in optics, photonics, and radiophysics.
\end{abstract}

\newpage 


\section{Introduction}
\begin{figure}[t]
\includegraphics[width=0.5\columnwidth]{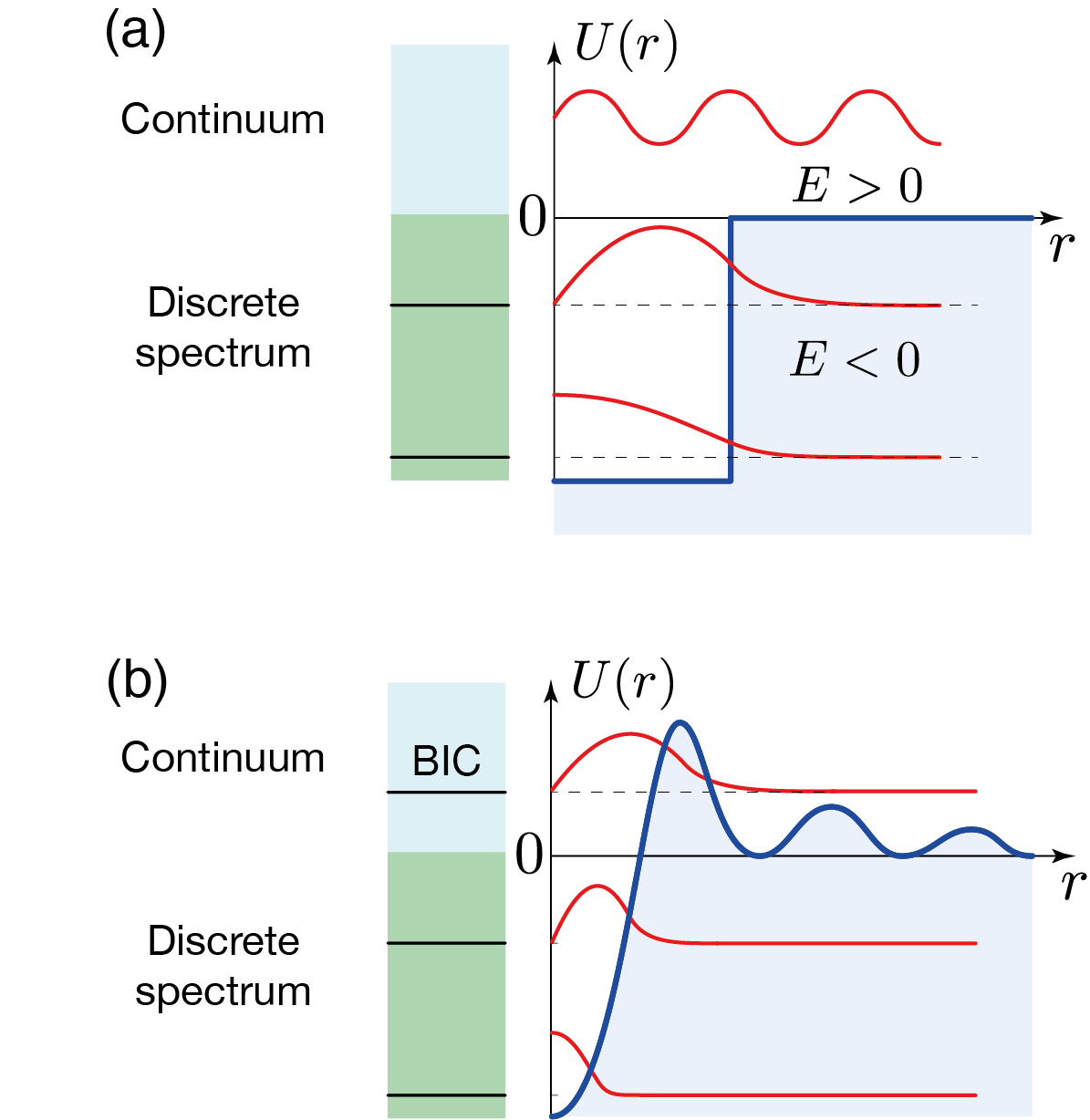}
\caption{(a) Spherical quantum well: the spectrum is continuous for $E>0$ and discrete for $E<0$; (b) 
\textcolor{black}{A specific potential results in the formation of} bound states 
\textcolor{black}{in the} continuum} \label{fig:1a1}
\end{figure}

One of the basic problems in quantum mechanics is the \textcolor{black}{energy} eigenvalue problem of a particle in a spherical quantum well.  Below the barrier ($E<0$), the spectrum is discrete and the wavefunctions $\psi(\mathbf{r})$ are bounded, i.e. $\int_{\mathcal{R}^3}\psi(\mathbf{r})d\mathbf{r}<\infty$.
Above the barrier ($E>0$), the spectrum is continuous and the wavefunctions
\textcolor{black}{cannot be normalized in the classical sense} [see Fig.~\ref{fig:1a1}(a)].
These solutions can be presented as propagating modes of the free space surrounding the quantum well. However, E. Wigner and J. von Neumann revealed in 1929 that this 
\textcolor{black}{classification} can be broken for specific potentials that asymptotically tend to zero away from the quantum well~\cite{Neumann2}. They 
\textcolor{black}{showed} that for some potentials, there are bound states embedded in the continuum of propagating modes [see Fig.~\ref{fig:1a1}(b)]. \textcolor{black}{Today,} such modes \textcolor{black}{are} 
known as {\it bound states in the continuum} (BICs). It seems that this term was introduced by L. Fonda in 1960~\cite{fonda1960theory}.  We should mention that the 
work \textcolor{black}{by} E. Wigner and J. von Neumann contains an algebraic error that was mentioned and corrected \textcolor{black}{by Stillenger and Herrick} in Ref.~\cite{stillinger1975bound}. \textcolor{black}{Later,} the theory of BICs was extended for the description of 
\textcolor{black}{ various} atomic, molecular, and quantum mechanical systems~\cite{stillinger1974role,stillinger1975bound,robnik1986simple,pappademos1993bound,friedrich1985physical,nockel1992resonances,cederbaum2003conical,sadreev2006bound}, but the potential proposed by E. Wigner and J. von Neumann is rather specific and it has been never implemented. Some ideas \textcolor{black}{on} how to construct potentials supporting BIC in semiconductor superlattices were developed in Refs.~\cite{herrick1976construction,stillinger1976potentials}.
\textcolor{black}{However,} they also were not implemented experimentally. We should mention that the experiment in \cite{capasso1992observation} has nothing \textcolor{black}{in} common with the observation of BICs in semiconductor superlattices. The Authors \textcolor{black}{of this work simply}
observed a defect state in the bandgap spatially localized by \textcolor{black}{electron Bragg reflectors.}
BICs are not \textcolor{black}{a} unique feature of quantum mechanical systems.
\textcolor{black}{In contrast, they are} specific solutions of 
wave equations in general 
\textcolor{black}{and} they can exist in acoustics, hydrodynamics, aerodynamics~\cite{parker1967resonance, parker1966resonance, evans1994existence, ursell1991trapped, ursell1951trapping, jones1953eigenvalues,lyapina2015bound}, and optics~\cite{marinica2008bound,bulgakov2008bound}. In acoustics and hydrodynamics, 
BICs 
\textcolor{black}{have been} known for a long time as {\it trapped modes}~\cite{jones1953eigenvalues,Callan1991Aug,Pagneux2013,Cobelli2011Jan}.

In recent years, \textcolor{black}{BICs are actively studied in the areas of }
optics and photonics, \textcolor{black}{as these states open up} 
tremendous opportunities for \textcolor{black}{implementing} 
compact high-Q resonators and metasurfaces required for biosensing and enhancement \textcolor{black}{of} nonlinear optical effects and light-matter interaction. \textcolor{black}{In the last few years, several} 
reviews of the state-of-the-art achievements 
in BICs were published 
~\cite{hsu2016bound,sadreev2021interference,azzam2021photonic,koshelev2019meta,koshelev2020engineering} . In this review, we focus 
on BICs in electrodynamic
systems \textcolor{black}{in various spectral ranges,} including visible, infrared, terahertz, and microwave
\textcolor{black}{ ones. In this review, we present} history, modern achievement\textcolor{black}{s}, and different physical 
models explaining the nature of this beautiful phenomen\textcolor{black}{on. }
\textcolor{black}{In the first section, the history of BICs in optics is described, and a selection of pioneering scientific papers is presented, including those that were published before the introduction of the term ‘optical bound states in the continuum’. The second section is devoted to the description of various mechanisms of BICs formation in dielectric structures. Their quantum-mechanical nature, the importance and role of periodic potential for BICs implementation, and topological properties are discussed; an explanation of BICs properties in terms of multipole analysis is also provided. In the third section, examples of photonic structures with BICs are presented, and features of BICs are described in structures of various dimensionality: from single nanoparticles to periodic metasurfaces. In this section, we also briefly talk about quasi-BICs, which are formed from BICs due to the violation of symmetry of a photonic structure. In the fourth section, the applications of BICs are discussed for the detection of biological objects, laser generation, twisted light beams, and optical harmonics.}

\section{Historical reference} 
\label{sec:history}
 
\begin{figure}[t]
\centering
\includegraphics[width=0.98\linewidth]{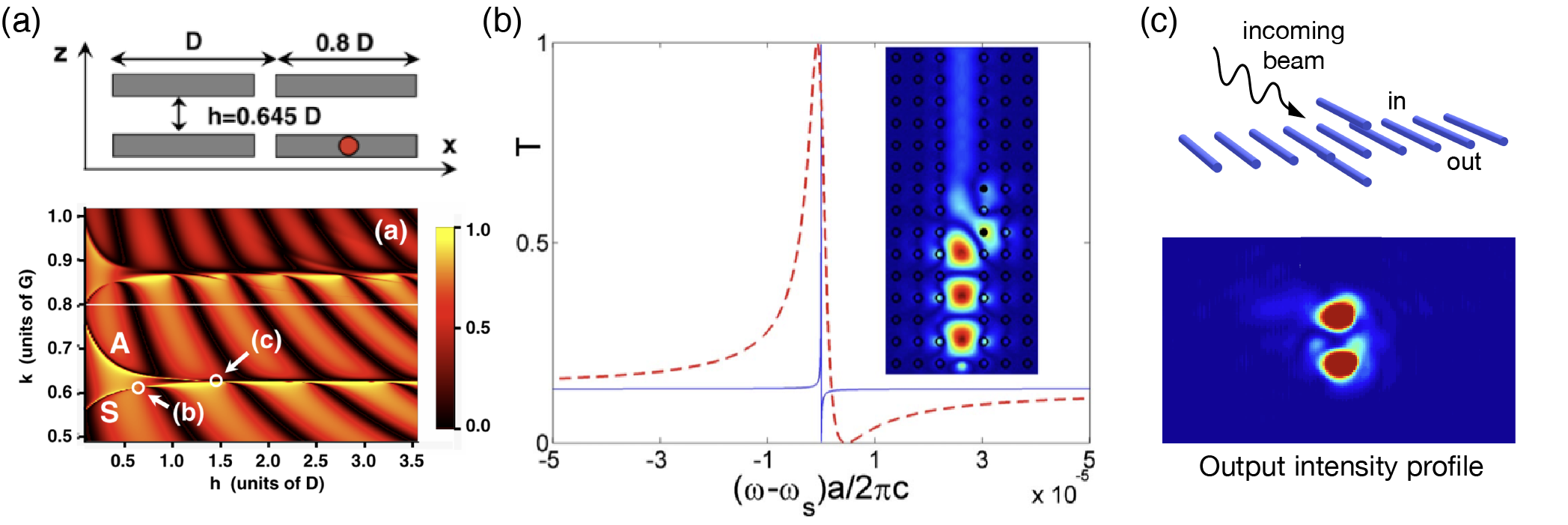}
\caption{(a) Upper: schematic of \textcolor{black}{a} double grating structure in vacuum. Lower: specular reflection coefficient as a function of the wave vector $k$ of the incident radiation and the distance $h$ between the gratings for the fixed value of $k_{x}=0.2G$. Adapted from Marinica et al.~\cite{marinica2008bound}. (b) The probability of transmission of \textcolor{black}{a} propagating photonic mode as 
\textcolor{black}{a function of} the frequency in the vicinity of \textcolor{black}{the} BIC for two sets of \textcolor{black}{permittivity values. }
\textcolor{black}{Inset:} the mode profile. 
Adapted from Bulgakov et al.~\cite{bulgakov2008bound}. (c) Upper: schematic of 
\textcolor{black}{a} one-dimensional array of $51$ 
and two additional waveguides above and below the array. Lower: light intensity at the output plane of the structure. Adapted from Plotnik et al.~\cite{plotnik2011experimental}.} \label{fig:1d1}
\end{figure}

\textcolor{black}{In modern literature, it is widely assumed that}
BICs in 
\textcolor{black}{optical systems were predicted} in 2008 in two works published by Marinica, Borisov, and \textcolor{black}{Shabanov}~\cite{marinica2008bound},  and by Bulgakov and Sadreev~\cite{bulgakov2008bound}, 
\textcolor{black}{and that} the first experimental study of BICs in optics was performed in 2011 \textcolor{black}{in the work}~\cite{plotnik2011experimental},
\textcolor{black}{  where} Marinica 
\textcolor{black}{et al.} considered two examples of similar periodic photonic structures supporting 
BICs. One of these structures is shown in Fig.~\ref{fig:1d1}(a). This is 
\textcolor{black}{a} two-layer dielectric grating periodic along the $x$-\textcolor{black}{axis,} 
 and it has 
 translation\textcolor{black}{al} symmetry along the $y$-axis. The Authors show\textcolor{black}{ed} that at certain distances between the gratings, the resonances in the 
 \textcolor{black}{reflection spectrum} at oblique incidence become infinitely narrow, i.e. disappear from the spectra. They also mention\textcolor{black}{ed} that 
 \textcolor{black}{this state} corresponds to the light perfectly 
 \textcolor{black}{propagating} along the structure with 
 \textcolor{black}{no} radiation losses. Bulgakov and Sadreev  \textcolor{black}{also} consider\textcolor{black}{ed} a single-mode waveguide formed by two \textcolor{black}{identical} photonic crystals. They 
 show\textcolor{black}{ed} that light can be perfectly trapped at the defects of the photonic crystal despite the \textcolor{black}{fact that the} frequency of \textcolor{black}{the} trapped mode lies in the transmission band of the waveguide. Figure~\ref{fig:1d1}(b) shows the transmission spectr\textcolor{black}{a} 
 through the waveguide for different parameters of the defects (solid and dashed curves). The solid curve correspond\textcolor{black}{s} to the near-BIC case when Fano resonance collapses. The inset show\textcolor{black}{s} the field distribution at the minimum of the transmission. 

Plotnik and colleagues performed a beautiful experiment on observation of the symmetry-protected BICs in \textcolor{black}{a} photonic structure 
\textcolor{black}{that is} an array of parallel dielectric single-mode waveguides fabricated 
\textcolor{black}{of} fused silica by 
direct laser writing [Fig.~\ref{fig:1d1}(c)]~\cite{plotnik2011experimental}. The \textcolor{black}{near-field} coupling between the waveguide\textcolor{black}{s} results in the formation of 
\textcolor{black}{a} transmission band. Two additional waveguides fabricated above and below the array support the anti-symmetric mode with \textcolor{black}{a} frequency lying in the transmission band of the \textcolor{black}{waveguide} array. 
This anti-symmetric mode 
\textcolor{black}{was excited} from one side of the sample\textcolor{black}{, and the intensity distribution was} observed on the other \textcolor{black}{side}. 
\textcolor{black}{The} lower panel in Fig.~\ref{fig:1d1}(c) shows that the \textcolor{black}{energy of the} initial anti-symmetric mode does not 
\textcolor{black}{leak to the waveguide array.} In order to 
break the vertical symmetry, 
\textcolor{black}{a gradient of the} refractive index 
\textcolor{black}{was created} by heating the top side of the sample while cooling the bottom. Such heating results in the coupling of the excited \textcolor{black}{anti-symmetric} mode to the modes of the array.       
\textcolor{black}{Thus, as the mode propagates through the sample, its energy is distributed between the waveguides forming the array.} We would like to highlight that these three works 
revealed \textcolor{black}{only} the explicit connection between BICs in quantum mechanics and \textcolor{black}{BICs} optics for the first time. However, there are many earlier works
where \textcolor{black}{electromagnetic} BICs were studied theoretically and 
experimentally but 
\textcolor{black}{were not associated with} quantum mechanics and \textcolor{black}{the} pioneering work by E. Wigner and J. von Neumann.

\begin{figure}[t]
\includegraphics[width=0.95\columnwidth]{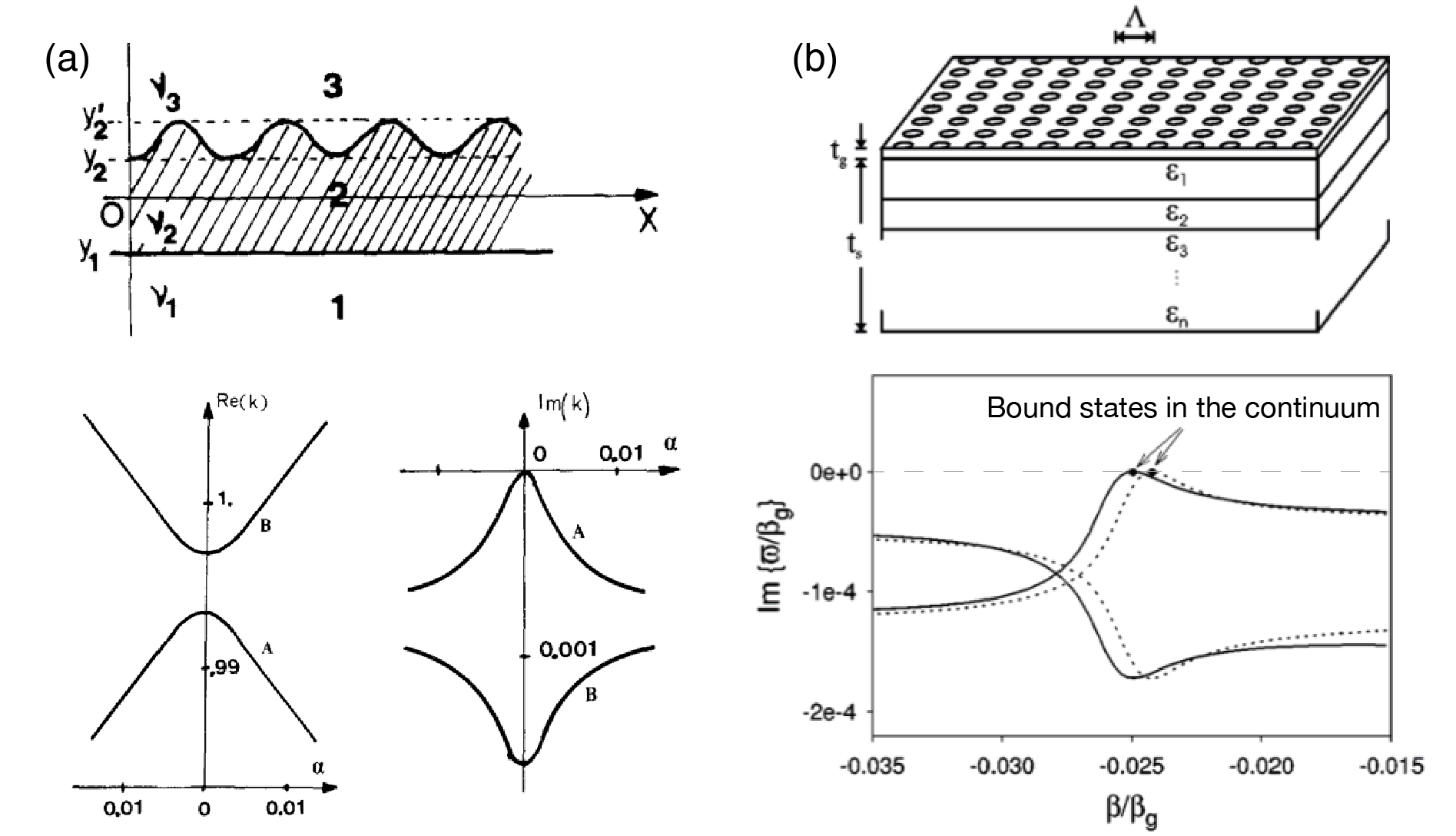}
\caption{(a) Upper: Schematic representation of a corrugated waveguide. Lower: Dispersion curves \textcolor{black}{for} 
the normalized frequency. 
The curves \textcolor{black}{correspond to the} two distinct branches: the anti-symmetric one (A) and the symmetric 
one (B). Adapted from 
\textcolor{black}{the work by Vincent and Neviere}~\cite{vincent1979corrugated} (b) Upper: schematic of a 
multilayer slab waveguide \textcolor{black}{with two-dimensional periodic texture}. Lower: Imaginary part of frequency 
near the anti-crossing between TE-like and TM-like \textcolor{black}{modes. }
The imaginary part of frequency \textcolor{black}{turns} 
to zero for the higher-energy band-edge states near the anti-crossing, indicating a truly bound 
state. Adapted from Paddon et al.~\cite{paddon2000}}. \label{fig:1b2}
\end{figure}

To the best of our knowledge, the history of BICs in optics began in 1976 with the work of Kazarinov, Sokolova, and Suris~\cite{kazarinov1976planar} who  consider\textcolor{black}{ed} a corrugated waveguide playing 
\textcolor{black}{the} role of 
\textcolor{black}{a} distributed\textcolor{black}{-}feedback resonator of 
\textcolor{black}{a} semiconductor laser. The Authors mentioned that if the eigenmode in the center of the Brillouin zone is formed by two counter-propagating waves which are $\pi$ out of phase, then the radiation losses are canceled. Independently from this work, Vincent and Neviere considered theoretically the band structure of a dielectric corrugated waveguide [see Fig.~\ref{fig:1b2}(a)]~\cite{Vincent1979}. The\textcolor{black}{y} showed that some resonances at the $\Gamma$-point of k-space are completely decoupled from the radiation continuum due to the symmetry mismatch \textcolor{black}{between the mode and the external field, which leads to} 
an infinite radiative lifetime \textcolor{black}{of the mode}. The authors of these works 
\textcolor{black}{did not associate} the observed resonances 
\textcolor{black}{with} BICs 
\textcolor{black}{and they did not} recognize the connection with earlier works in quantum mechanics, \textcolor{black}{but the physics of non-radiating states were studied in detail.}

\begin{figure}[t]
\includegraphics[width=0.95\columnwidth]{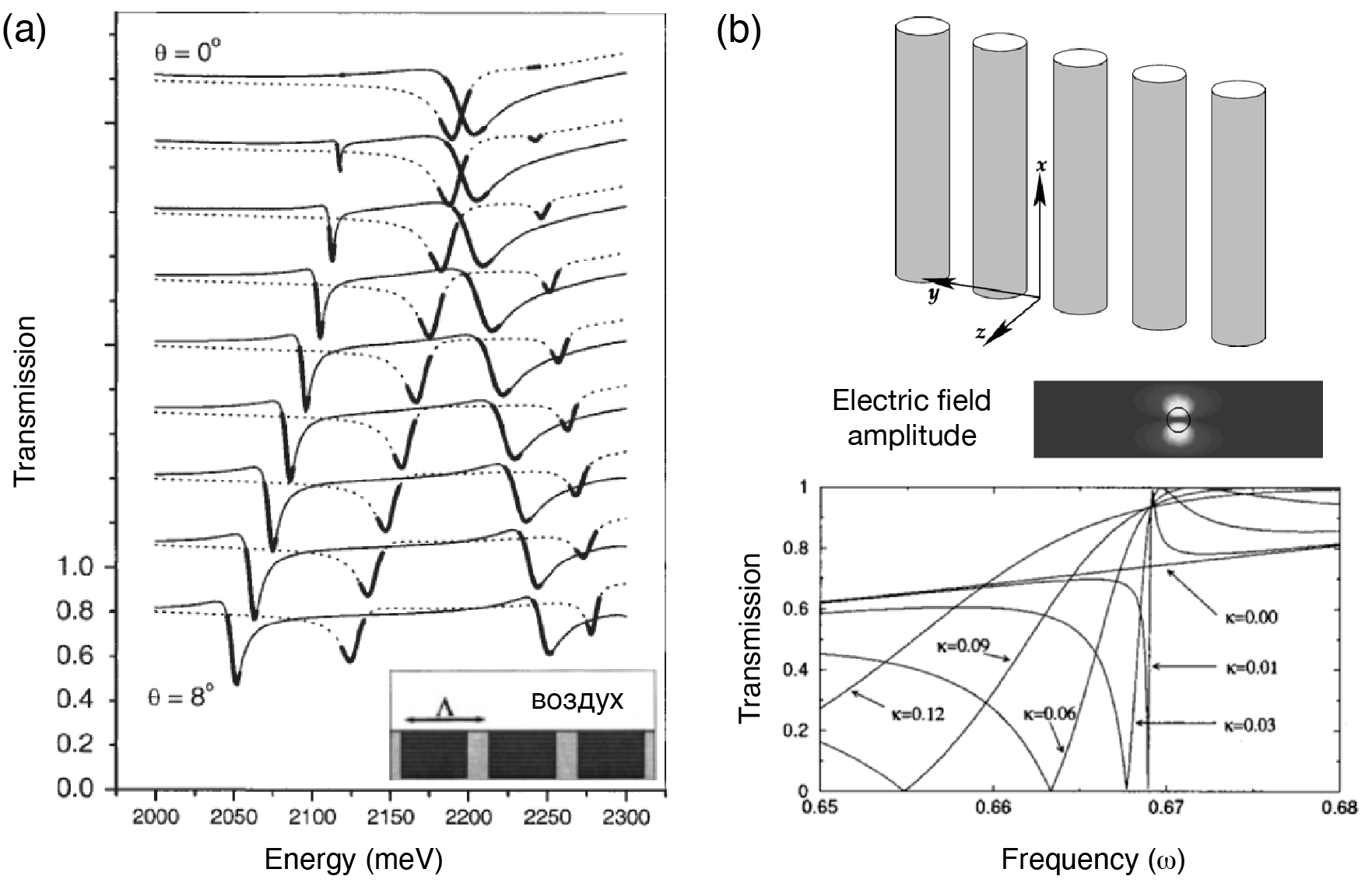}
\caption{(a) The transmission spectra of \textcolor{black}{a} polaritonic crystal slab shown in the inset. \textcolor{black}{The spectra are} calculated for different \textcolor{black}{angles of} incidence, in \textcolor{black}{the} narrow energy interval around the fourth Bragg resonance of \textcolor{black}{the} lower polaritonic branch.  line: 
\textcolor{black}{the incident wave is polarized} along the grooves (s-polarisation). Dotted line: \textcolor{black}{the incident wave is polarized ortogonal to} 
the grooves (p-polarisation). Adapted from Yablonskii et al.~\cite{yablonskii2002optical} (b) Upper panel: schematic of a two-dimensional \textcolor{black}{multilayer} periodic slab. Middle panel: 
amplitude of \textcolor{black}{the field of} the bound state at $\kappa = 0$. Lower panel: \textcolor{black}{Results of the} numerical simulation \textcolor{black}{using} boundary integral equations of transmission vs 
\textcolor{black}{normalized frequency} for 
\textcolor{black}{TE-}polarized plane wave 
\textcolor{black}{incident on} a slab of vertical rods in air for various values of \textcolor{black}{y component of} the wave \textcolor{black}{vector.}
The \textcolor{black}{permittivity of the cylinders} 
is $12$, and the magnetic permeability is $1$. Adapted from Shipman et al.~\cite{shipman2005resonant}} \label{fig:3a1}
\end{figure}

\textcolor{black}{Nonradiative states were also found in a two-dimensional periodic array of coupled dielectric spheres by M.~Inoue and colleagues~\cite{inoue1982light}. They showed that at the center of the Brillouin zone, optically inactive states exist that cannot be excited by a normally incident wave, regardless of its polarization. At the same time, they can be excited at oblique incidence, and in this case, extremely narrow peaks appear in the reflection/transmission spectrum, see Fig.~\ref{fig:3a1}(a). The existence of non-radiating modes in photonic crystal structures was also discussed by K.~Sakoda~\cite{sakoda1995optical,sakoda1995transmittance} and later by P.~Paddon and J.~Young~\cite{paddon2000}. The latter have developed a method based on Green's functions that allows analyzing a complex photonic band structure, i.e. the spectral positions of resonances and their radiative lifetimes [see Fig.~\ref{fig:1b2}(b)].
This is probably the first work where BICs with a non-zero Bloch wavenumber, the so-called parametric (tunable) BICs or accidental BICs, were theoretically predicted. The authors explained that the coupling between the TE and TM modes leads to the anti-crossing of their dispersion curves and the formation of a mode with zero imaginary part of its eigenfrequency, i.e., the formation of BIC.
In fact, the described mechanism is identical to that analyzed in the work of H.~Friedrich and D.~Wintgen for quantum mechanical systems~\cite{friedrich1985physical}. Therefore, tunable BICs are also called Friedrich-Wintgen BICs. In 2003, S.~Shipman and S.~Venakides also observed BICs numerically in the transmission spectra through an array of parallel dielectric cylinders [Fig.~\ref{fig:3a1}(b)]. In the same year, they developed a theory explaining the formation of bound states and anomalies in the transmission spectrum corresponding to these states~\cite{shipman2005resonant}. Figure \ref{fig:3a1}(b) (bottom panel) shows the numerical transmission spectrum for various Bloch wave numbers. It is explicitly shown that at normal incidence, a collapse of the Fano resonance occurs. The inset in the middle part of the figure shows the electric field distribution of the bound state. In fact, it is anti-symmetric with respect to the plane of symmetry of the unit cell; however, the authors plotted the distribution amplitude, which is an even function. In 2003, Bonnet et al~\cite{bonnet2003high} discovered ultra-narrowband resonant reflection from a one-dimensional corrugated waveguide at oblique incidence, i.e. at non-zero Bloch wave number, and analyzed the effect of a finite beam size on the reflection spectrum.}

It seems that the \textcolor{black}{BICs were observed in experiment for the first time}
in the work by Henry et al 
in 1985~\cite{henry1985observation}. The Authors consider\textcolor{black}{ed} 
distributed-feedback resonator \textcolor{black}{with} 
a second-order grating. They 
showed that the losses of the lasing mode 
mainly occur at the ends of the structure, while 
at its center \textcolor{black}{they nearly vanish} due to \textcolor{black}{the} destructive interference of the scattered radiation from the 
\textcolor{black}{counter-}propagating waves forming the lasing mode, \textcolor{black}{as it was predicted in the work~\cite{kazarinov1976planar}}. Later on, in 1986~\cite{Sychugov1986}, Avrutskii 
\textcolor{black}{et al.} analyzed the reflectance spectra from the corrugated ZnO waveguide deposited on a glass substrate in the visible range. It was found that 
\textcolor{black}{a} peak \textcolor{black}{in }
the second stop band disappear\textcolor{black}{s} from the spectrum at the normal incidence, manifesting the bound state form\textcolor{black}{ed by }
counter-propagating waveguide modes.       

Another experiment was done by Robertson and coauthors in 1992~\cite{robertson1992measurement}. They analyzed the transmission spectra of \textcolor{black}{a} two-dimensional dielectric structure consisting of alumina-ceramic cylinders arranged in a square array. \textcolor{black}{The experiment was performed} in the GHz frequency range (10-150 GHz). The Authors mentioned that one of the band\textcolor{black}{s was} 
not observed in the experiment \textcolor{black}{because of the symmetry mismatch between the eigenmode and the exciting field}. 
An experiment on the  observation of BIC in the visible range was made in \cite{Pacradouni2000} by Pacradouni and colleagues. They measured the \textcolor{black}{reflection} 
spectra with angular resolution from the perforated AlGaAs membrane, confirming the 
\textcolor{black}{narrowing} lines in the vicinity of \textcolor{black}{both} at-$\Gamma$ BIC (at normal incidence) and off-$\Gamma$ BIC (at oblique incidence). However, we should \textcolor{black}{note} 
that this work contains a minor inaccuracy. 
\textcolor{black}{Specifically, the existence of BICs with non-zero Bloch wave vector} requires up-down mirror symmetry~\cite{zhen2014topological} or \textcolor{black}{a} very fine adjustment of the \textcolor{black}{geometric and material parameters of the system}
~\cite{bulgakov2018propagating}, which were not performed, according to the article. Thus, 
Pacradouni and colleagues 
observed \textcolor{black}{only} the increase of the Q factor, but not a BIC.
For observation of off-$\Gamma$ \textcolor{black}{BIC, i.e. with a non-zero Bloch vector,} in photonic crystal membranes, the structure is usually suspended~\cite{jin2019topologically} 
\textcolor{black}{or covered by an optical liquid} 
\textcolor{black}{whose} refractive index \textcolor{black}{matches that} of the substrate~\cite{hsu2013observation}, \textcolor{black}{which provides the up-down mirror symmetry}. The at-$\Gamma$ BIC was also observed experimentally 
\textcolor{black}{in a} polaritonic system in 1998, by Fujita and colleagues~\cite{fujita1998tunable}. The Authors analyzed the angular-resolved transmission spectra of 
\textcolor{black}{a} distributed-feedback microcavity consisting of quartz grating substrate covered by 
\textcolor{black}{an organic-inorganic} perovskite-type semiconductor. 
\textcolor{black}{They discovered} that at normal incidence 
the resonances disappear. These experimental results were comprehensively described by Yablonskii and colleagues in~\cite{yablonskii2001polariton}.

In this section, we trie\textcolor{black}{d to review the} 
key works 
\textcolor{black}{on BICs} in electromagnetic system\textcolor{black}{s}. However, we 
\textcolor{black}{should note} that the existence of non-radiating states in photonic structures \textcolor{black}{was} also discussed in many other works (see, for example, Refs.~\cite{Fan2002,tikhodeev2002quasiguided,ochiai2001dispersion,Sychugov1986}). To conclude this historical \textcolor{black}{summary}, 
we should mention 
that the physics of non-radiating states in periodic structures is quite clear, and 
\textcolor{black}{other} earlier work\textcolor{black}{s might exist} where \textcolor{black}{such} 
states are discussed.

\section{BICs in dielectric photonic structures}

\subsection{From quantum mechanics to photonics}

\begin{figure}[t]
\includegraphics[width=0.7\columnwidth]{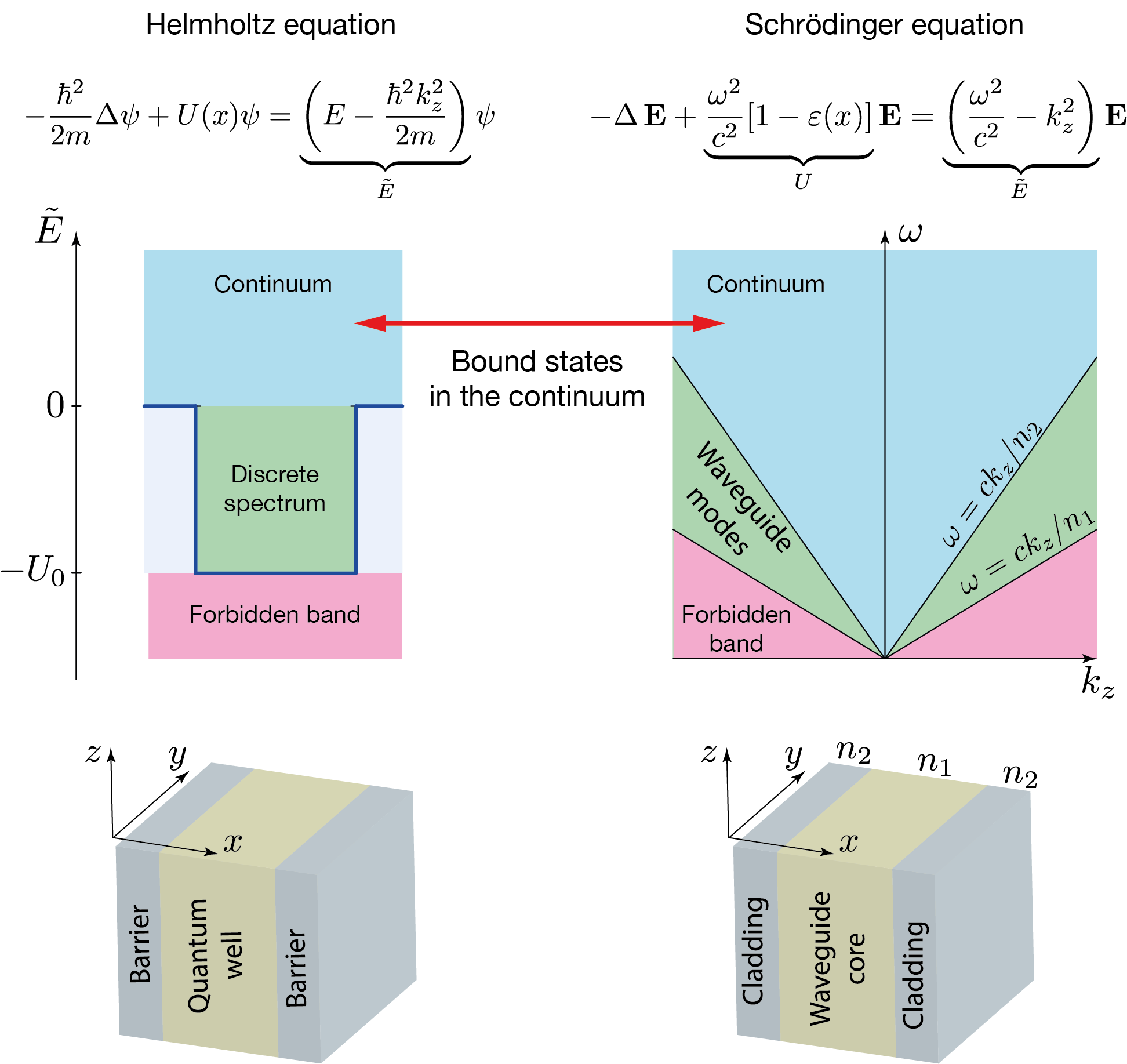}
\caption{Correspondence between quantum mechanical and electromagnetic problems \textcolor{black}{for an 1D potential} having a translational symmetry along \textcolor{black}{the} $z$-direction. Upper panels: Schrödinger equation for 1D quantum well, and Helmholtz equation for a plane dielectric waveguide. Middle panels: potential well (left) and the dispersion diagram (right). Lower panels: potential well (left) and planar waveguide (right) in the coordinate space.} \label{fig:1b1}
\end{figure}

Let us set a link between quantum mechanics and optics by the example of a 1D quantum well with a translation\textcolor{black}{al} symmetry along the $z$-direction. For such \textcolor{black}{a} system, we can define the domain of the continuum spectrum as $\tilde E = E -\frac{\hbar^2 k_z^2}{2m}>0$ \textcolor{black}{[see Figure~\ref{fig:1b1}, left panel]}. An optical analogue of 
\textcolor{black}{this} system is a parallel-plate dielectric waveguide. By rewriting the Helmholtz equation to the form of the stationary Schr\"{o}dinger equation \textcolor{black}{[see Figure~\ref{fig:1b1}, right panel]}, one can see that the 
permittivity $\varepsilon(x)$ can be associated with quantum mechanical potential $U(x)$. The waveguide modes lying under the light line $\omega<ck_z$ represent 
discrete states and all \textcolor{black}{the} modes with $\omega>ck_z$ form 
\textcolor{black}{the} continuum. Thus, in the system with a translation\textcolor{black}{al} symmetry or periodicity along 
\textcolor{black}{a certain} direction, BICs will be localized only in the orthogonal directions. The analogy between quantum and optical systems is very 
\textcolor{black}{illustrative, but it is} not complete. The vector structure of electromagnetic field\textcolor{black}{s} (polarization) makes the electromagnetic systems more diverse. \textcolor{black}{We should also note that the Helmholtz equation written in the form of the stationary Schrödinger equation (Fig.~\ref{fig:1b1}) is a so-called generalized eigenvalue problem, since the required frequency $\omega$ is included both in the eigenvalue (the right side of the equation) and in the potential. However, if we consider the wave vector $k_z$ as an eigenvalue, then we obtain a classical eigenvalue problem.}

\begin{figure}[t]
    \centering
    \includegraphics[width =0.6\linewidth] {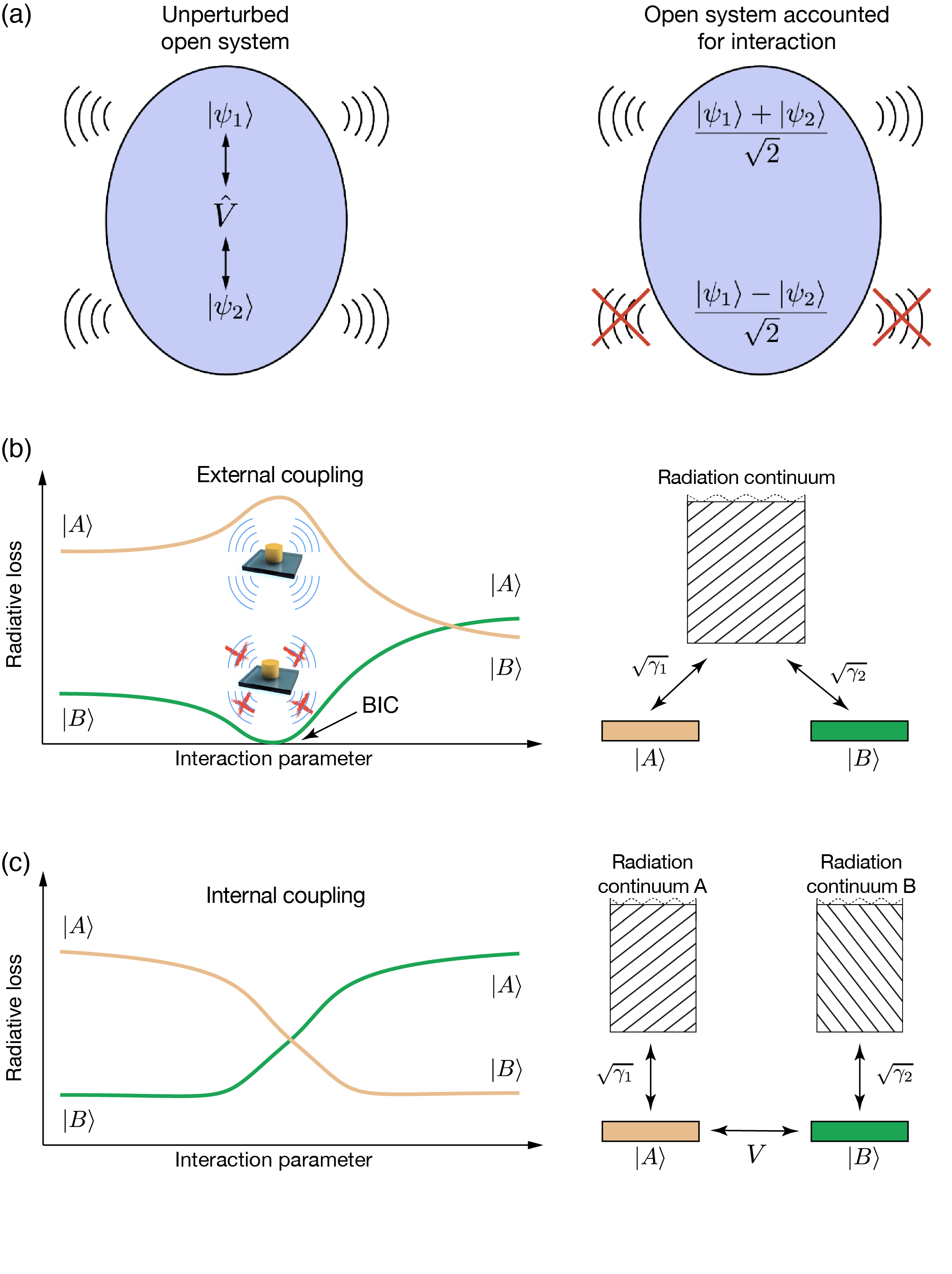}
    \caption{Illustration of \textcolor{black}{the} 
    general principle explaining BICs proposed by Fri\textcolor{black}{e}drich and Wintgen: (a) nonperturbed and perturbed open systems; (b) open system with two interacting leaky modes coupled to the same 
    continuum, which allows for an existence of BIC; (c) open system with two interacting leaky modes coupled to different radiative continua, no BIC is allowed.}
    \label{fig:2a1}
\end{figure}

One of the general principle\textcolor{black}{s} that can explain the appearance of BICs in most electromagnetic systems is the destructive interference of two interacting leaky waves. This mechanism was originally proposed by Fri\textcolor{black}{e}drich and Wintgen~\cite{friedrich1985physical}, and it is schematically shown in Fig.~\ref{fig:2a1}(a). Let an open system (resonator) has two leaky modes $\ket{\psi_s}$ ($s=1,2$) with close \textcolor{black}{or even equal } 
eigenfrequencies $\Omega_s=\omega_s+i\gamma_s$ ($s=1,2$). Then, let\textcolor{black}{'s} introduce some perturbation $\hat V$ that makes these states coupled. In the framework of the perturbation theory, the eigenmodes in the system, with account for the interaction, 
\textcolor{black}{can be presented as} a linear superposition of the initial states $\ket{\psi}=C_1\ket{\psi_1}+C_2\ket{\psi_2}$. If we can continuously tune the interaction potential $\hat V$, then at some 
specific conditions, the radiative losses can be suppressed completely (substantially). In this case, a genuine 
BIC (\textcolor{black}{or a} quasi-BIC) appears. 

More formally, an open system within \textcolor{black}{the} two-mode approximation can be described in the framework of the temporal coupled-mode theory, where $\mathbf{a}=[a_1(t), a_1(t)]^T$ are the complex amplitudes of states  $\ket{\psi_1}$ and $\ket{\psi_1}$. The complex amplitudes evolve in time as: 
\begin{align}
\label{eq:CMT1}
& \frac{d\mathbf{a}}{dt}=\hat H \mathbf{a},\\
&\hat H=
\underbrace{
\begin{pmatrix}
\omega_{1}      & \kappa  \\
\kappa          & \omega_{2}
\end{pmatrix}}_{\hat H_0}
-i
\underbrace{\begin{pmatrix}
\gamma_1                                    & \sqrt{\gamma_{1} \gamma_{2}}e^{i\phi} \\
\sqrt{\gamma_{1} \gamma_{2}}e^{i\phi}       & \gamma_2
\end{pmatrix}}_{\hat V}.
\label{eq:CMT2}
\end{align}
Here, $\kappa$ is responsible for the internal coupling, $\sqrt{\gamma_1\gamma_2}$ accounts for the coupling through the radiation continuum, and $\phi$ is the
phase shift between the modes. The condition of the BIC appearance in \textcolor{black}{the} two-mode approximation can be written as~\cite{volya2003non}
\begin{align}
&\kappa(\gamma_1-\gamma_2)=e^{i\phi}\sqrt{\gamma_1\gamma_2}(\omega_1-\omega_2) \\     
& \phi=\pi m, \ \text{where} \ m\in\mathcal{Z}
\end{align}
These conditions can be fulfilled through the tuning of the parameters of two coupled resonances. Figure~\ref{fig:2a1}(b) shows schematically how the radiative losses of \textcolor{black}{the} resonant states $\ket{\psi_1}$ and $\ket{\psi_2}$ depend on the coupling strength $\kappa$. It is worth mentioning that exactly at the point where the BIC appears, the radiative losses for the second mode are equal to $\gamma_1+\gamma_2$. This is  an analogue of the Dicke superradiance for two emitters~\cite{dicke1954coherence,mlynek2014observation}. 

In the framework of this simple model, the initial resonant state\textcolor{black}{s} should radiate to the same radiative continuum (\textcolor{black}{scattering channel}). Only in this case they can interfere destructively and form a BIC. If the states $\ket{\psi_1}$ and $\ket{\psi_2}$ radiate to different radiative continua [see Fig.~\ref{fig:2a1}(c)], then the radiative losses of the dressed states will be always between $\gamma_1$ and $\gamma_2$~\cite{cao2015dielectric}.     

\begin{figure}[t]
    \centering
    \includegraphics[width =0.6\linewidth] {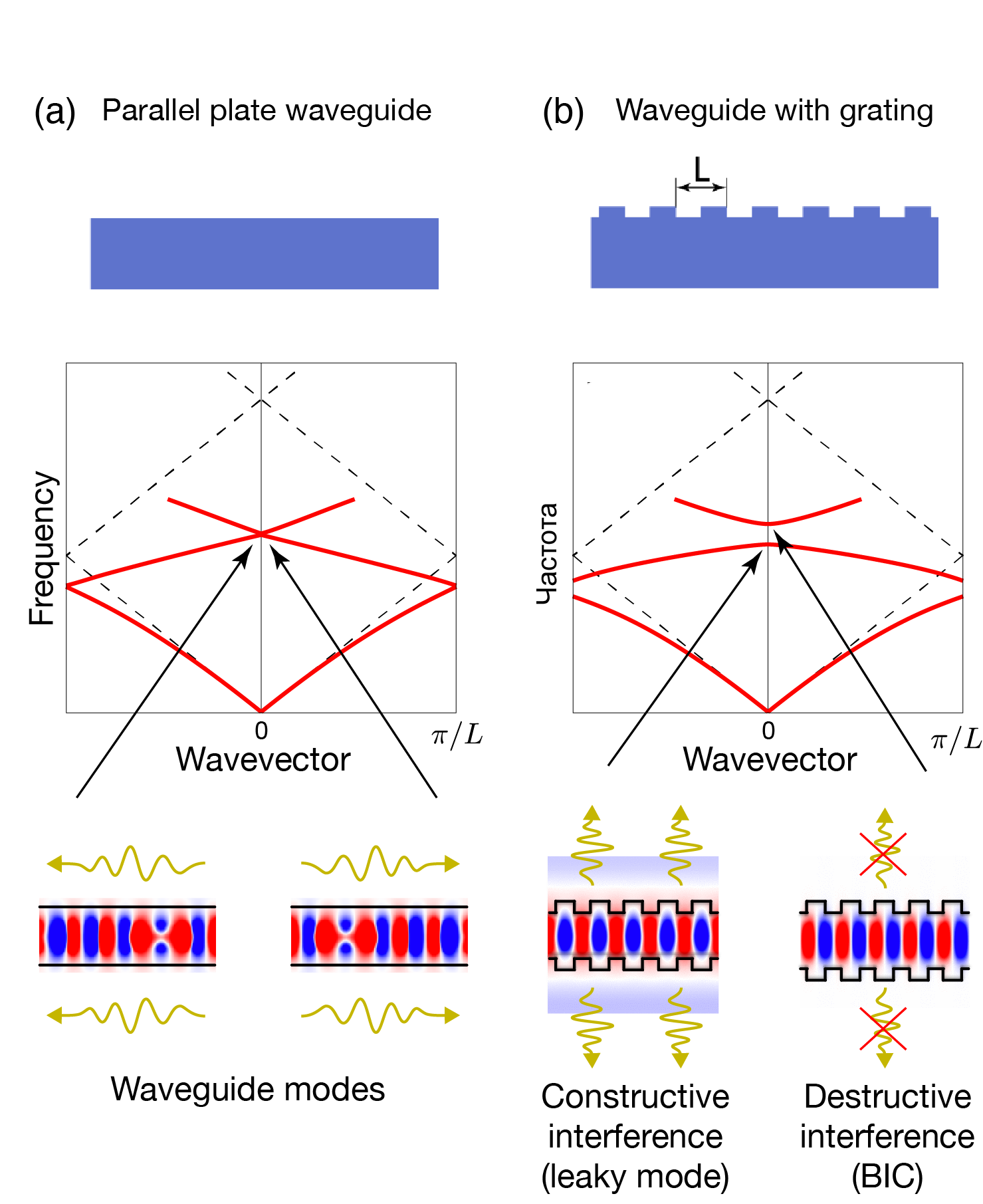}
    \caption{(a) Geometry, folded band structure and mode profiles for a dielectric waveguide. The waveguide supports guided modes propagating to the left and right. (b) Geometry, folded band structure and mode profiles for a modulated dielectric waveguide with period $L$. Due to \textcolor{black}{the} periodicity, the waveguide modes interact and form two new states due to constructive (leaky mode) and destructive (BIC) interference. }
    \label{fig:2a2}
\end{figure}

Now, let us illustrate how this model 
\textcolor{black}{explains the formation of BICs in} a corrugated dielectric waveguide. First, let's 
\textcolor{black}{consider a} periodic potential 
\textcolor{black}{with a vanishing} amplitude 
[see Fig.~\ref{fig:2a2}(a)]. \textcolor{black}{This is the so-called empty lattice approximation, which is well-known in solid state physics~\cite{kittel1996introduction}}. Therefore, the second 
\textcolor{black}{bandgap} is closed, and the eigenstates in the $\Gamma$-point are degenerate. They are counter-propagating leaky waves. Periodic potential with a finite amplitude lifts 
the degeneracy and opens the 
\textcolor{black}{bandgap.} If the periodic potential is symmetric with respect to the $z \rightarrow -z$ transformation, then the new s\textcolor{black}{t}ates at \textcolor{black}{the} edges of the 
\textcolor{black}{second bandgap are a BIC} 
(anti-symmetric combination of the leaky modes) and \textcolor{black}{a} superradiant state (symmetric combination of the leaky modes) [see Fig.~\ref{fig:2a2}(b)]. \textcolor{black}{This mechanism of BIC formation is quite clear, and it was discussed in many works, for example, in Refs. \cite{kazarinov1976planar,Vincent1979,gao2016formation,bulgakov2018avoided}.}

\subsection{BIC and diffraction orders}

A BIC can be formed from a leaky resonance \textcolor{black}{if the scattering amplitudes to all the channels turn to zero, i.e. in the case of} 
decoupling from 
\textcolor{black}{all the} open scattering channels. \textcolor{black}{This can be achieved} by varying parameters of the system. \textcolor{black}{In this case,} the BIC formation 
is possible only if the number of the adjust\textcolor{black}{ed}
parameters is \textcolor{black}{larger}  
than the number of the scattering channels. For finite-size structures, the number of scattering channels is infinite, and the existence of BICs in such systems is prohibited by the non-existence theorem~\cite{hsu2016bound}. Until recently, it was believed that the only exception is the structures surrounded by a completely opaque shell providing \textcolor{black}{the} decoupling of the internal resonances from the outside radiation continuum, which in quantum mechanics corresponds to an infinite 
potential barrier
, in acoustics to hard-wall boundaries, and in optics to perfect\textcolor{black}{ly} conducting walls or epsilon-near-zero barriers~\cite{monticone2014embedded,liberal2016nonradiating}. One more exception from the 'non-existence' theorem was found in~\cite{Deriy2022Feb}, where the authors revealed that finite-size solid acoustic resonators can support genuine BICs completely localized inside the resonator. Despite possible exceptions from the 'non-existence' theorem, the BIC\textcolor{black}{s are} 
usually \textcolor{black}{formed}
in structures with a finite number of scattering channels. A typical example of \textcolor{black}{such} a system 
is a resonator coupled to one or several waveguide modes~\cite{lepetit2014controlling,pilipchuk2020bound,lyapina2015bound} or infinite photonic structure\textcolor{black}{s} periodic \textcolor{black}{in one or two} 
directions~\cite{hsu2013observation,sadrieva_experimental_2019,zhen2014topological,marinica2008bound,yuan2020parametric}.
 
\begin{figure}[t]
    \centering
    \includegraphics[width =0.9\linewidth] {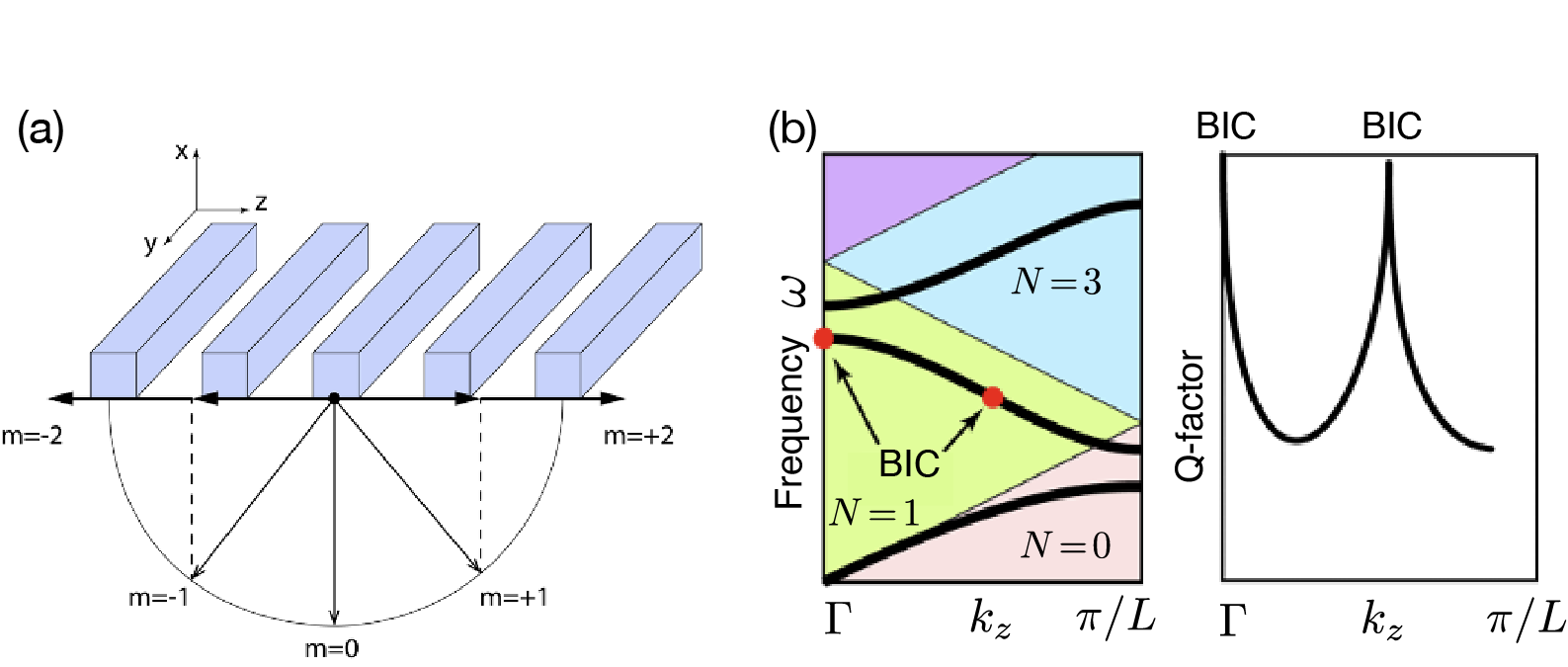}
    \caption{(a) Example of a 1D dielectric grating; (b) typical dispersion for 1D grating, \textcolor{black}{where colors indicate the domains with different }
    numbers of diffraction channels, and \textcolor{black}{the dots indicate the} positions of BICs (left); 
    variation \textcolor{black}{of} the Q factor along the second \textcolor{black}{dispersion curve}     
    (right).}
    \label{fig:2b1}
\end{figure}

 Now, let us \textcolor{black}{consider in more detail} 
 the mechanism of the BIC formation in periodic structures, by the example of \textcolor{black}{a} dielectric grating with \textcolor{black}{a} period $L$ shown in Fig.~\ref{fig:2b1}(a). The electric field, being a Bloch function, can be written as
\begin{equation}\label{eq:2_B_1}
\mathbf{E}_{n,k_b}(x,y,z)=e^{ik_bz+ik_yy}\mathbf{u}_{n,k_b}(x,z).
\end{equation}
Here, $k_b$ is the Bloch wavenumber, $k_y$ is the wavenumber component along the $y$-axis (the direction of the translation\textcolor{black}{al} symmetry), \textcolor{black}{and} $n$ is the index of the photonic band. 
Periodic function $\mathbf{u}_{n,k_b}(x,z)$ can be expanded into the Fourier series
\begin{equation}\label{eq:2_B_2}
\mathbf{u}_{n,k_b}(x,z)=\sum_{s} \mathbf{c}_{n,s}(x) e^{ik_bz+\frac{2\pi i s}{L}z}.
\end{equation}
Here, $s$ is an integer. Each term in this series 
correspond\textcolor{black}{s} to a diffraction channel, which could be open or closed. Outside the structure, the expansion coefficient\textcolor{black}{s} describe a plane wave as
\begin{align} \label{eq:2_B_2n}
&\mathbf{c}_{n,s}(x) \longrightarrow \mathbf{c}_{n,s}e^{\pm iK_s x}, \\
&K_s=\sqrt{\frac{\omega^2}{c^2}-k_y^2-\left(k_b^2+\frac{2\pi s}{L}\right)^2}.
\end{align}
If $K_s$ is real, then the diffraction channel is open and $\mathbf{c}_{n,s}$ is the complex amplitude \textcolor{black}{of the} wave outgoing via the $s$-th diffraction channel. If $K_s$ is imaginary, then the diffraction channel is closed and $\mathbf{c}_{n,s}$ is the complex amplitude of the near field.    

Figure~\ref{fig:2b1}(b) shows schematically the characteristic dispersion ($\omega$ vs $k_b$ for $k_y=0$) of eigenmodes in the dielectric grating. \textcolor{black}{The} color\textcolor{black}{ed} domains in the figure show the regions where a certain number of diffraction channels ($N$) are open. Under the light line, $N=0$. Thus, all \textcolor{black}{the} diffraction channels are closed, and only the waveguide modes exist in the structure. 

To construct a BIC, all \textcolor{black}{the} coefficients $\mathbf{c}_{n,s}$ corresponding to the open diffraction channels should be \textcolor{black}{equal to zero}. 
However, for subwavelength structure\textcolor{black}{s with} $L<\lambda$, there is only one open diffraction channel \textcolor{black}{($N=1$)} corresponding to $s=0$. Thus, 
\textcolor{black}{in order to form} a BIC, we need to nullify $\mathbf{c}_{0}(x)$. Here, we omit \textcolor{black}{the} index $n$ for the sake of simplicity.  \textcolor{black}{The} function $\mathbf{c}_{0}(x)$ is the zeroth Fourier coefficient in \textcolor{black}{the}  expansion~\eqref{eq:2_B_2}, Thus, according to the definition 
\begin{equation}
\mathbf{c}_{0}(x)=\int\limits_{-L/2}^{L/2} 
\mathbf{u}_{k_b}(x,z) e^{-ik_bz}dz.
\end{equation}
For the state in the center of Brillouin zone ($k_b=0$) one can write
\begin{equation}
\mathbf{c}_{0}(x)=\int\limits_{-L/2}^{L/2} 
\mathbf{u}_{k_b}(x,z)dz=\langle \mathbf{u}_{k_b}(x,z) \rangle_z. 
\end{equation}
Therefore, for BIC in the $\Gamma$-point, the $z$-averaged field should be equal to zero. If the unit cell of the grating is symmetric with respect to the $z\rightarrow-z$ transformation, then the eigenstates in the $\Gamma$-point can be even or odd functions of $z$. For odd functions, their  $z$-averaged value is zero. 
\textcolor{black}{Therefore,} all such state\textcolor{black}{s} are BICs. Th\textcolor{black}{ese}
BICs are called {\it symmetry-protected}, \textcolor{black}{which means} 
that the cancellation of radiation is protected by the symmetry of the structure. It is noteworthy that symmetry-protected BICs emerge in both low-contract and high-contrast photonic structures~\cite{bulgakov2018avoided,gao2019bound,wang2016optical}. As opposed to symmetry-protected BIC, there \textcolor{black}{are} 
\textcolor{black}{the so-called Friedrich-Wintgen BICs, also known as} {\it accidental} or {\it tunable} BIC\textcolor{black}{s,} for which $\mathbf{c}_{0}(x)$ nullifies not \textcolor{black}{because of} 
the symmetry reasons but due to the tuning of 
parameters of the system~\cite{hsu2013observation,zhen2014topological}.

\textcolor{black}{In the general case,} the Fourier amplitude $\mathbf{c}_{0}$ is a complex vector function \textcolor{black}{of the grating parameteres}. 
Thus, for BIC \textcolor{black}{formation}, both \textcolor{black}{its} real and imaginary parts 
should be zero. However, one can show that for BIC, the components of $\mathbf{c}_{0}$ can be chosen real everywhere in the $\mathbf{k}$-space if the structure possesses  time-reversal symmetry $\varepsilon^*(\mathbf{r})=\varepsilon(\mathbf{r})$, inversion symmetry $\varepsilon(\mathbf{-r})=\varepsilon(\mathbf{r})$, and up-down mirror symmetry~\cite{hsu2013observation,zhen2014topological}. However, \textcolor{black}{a} 
symmetry-protected BIC does not require up-down mirror symmetry, and \textcolor{black}{therefore it} can be implemented in structures with substrate. This symmetry is necessary for observation of a tunable (off-$\Gamma$) BIC. Thus, \textcolor{black}{as it was mentioned above, in experiments} the samples are usually immersed in an optical liquid that is index-matched to the substrate, or suspended structures are used~\cite{hsu2013observation,kodigala2017lasing}. 

\textcolor{black}{In order to create} a BIC, it is not necessary to work in a region 
\textcolor{black}{with only one open diffraction channel.} 
If several diffraction channels are open, one needs to tune several parameters of the system to 
\textcolor{black}{achieve a BIC. It is hard to implement in practice, but there are theoretical works showing it is possible~\cite{ndangali2010electromagnetic,bulgakov2014bloch}.}

\subsection{BICs and \textcolor{black}{the multipole expansion}}

The appearance of BICs in 
periodic photonic \textcolor{black}{structures}
can be explained in terms of \textcolor{black}{the multipole expansion}. 
The multipole approach is based on the formalism of the vector spherical harmonics (multipoles) representing a \textcolor{black}{complete} basis set of orthogonal vector functions that are the solutions of the vector Helmholtz equation~\cite{Jackson}. Usually, these function\textcolor{black}{s} are used in the Mie theory as independent scattering channels. They give
the connection between the directivity diagram of the outgoing radiation and \textcolor{black}{the} polarization currents induced inside a scatterer by \textcolor{black}{the} 
incident wave~\cite{Grahn2012}. A number of beautiful 
scattering phenomena \textcolor{black}{in the area of electromagnetic waves were} 
explained within the multipole approach including \textcolor{black}{the} anapole effect~\cite{miroshnichenko2015,yang2019nonradiating}, the Kerker effect~\cite{Poshakinskiy,shamkhi2018transverse,Liu2018}, \textcolor{black}{and} superscattering~\cite{ruan2010superscattering,ruan2011design,qian2019experimental,krasikov2021multipolar}. 
The multipole approach can \textcolor{black}{also enable a deeper understanding of} 
the physics of BICs in periodic structures~\cite{chen2019singularities,sadrieva2019multipolar}.

Following 
the book by Bohren and  Huffman\cite{Bohren}, we define the vector spherical harmonics $\mathbf{M}_{p \ell m}$ and $\mathbf{N}_{p \ell m}$ as

\begin{align}
    & \mathbf{M}_{p \ell m}=\nabla\times(\mathbf{r}\psi_{p \ell m}), \ \ \ p=o,e \\
    & \mathbf{N}_{p \ell m}=\frac{1}{k}\nabla\times\nabla\times(\mathbf{r}\psi_{p \ell m}), \ \ \ p=o,e \\
    &\psi_{ \left\{\begin{smallmatrix} e \\ o \end{smallmatrix} \right\} \ell m} = z_{\ell}(k r) P_{\ell}^{m}(\cos \theta) \left\{ \begin{array}{c} \cos m \varphi \\ \sin m \varphi \end{array} \right\}.
\end{align}

\begin{figure}[t]
    \centering
    \includegraphics[width =1\linewidth] {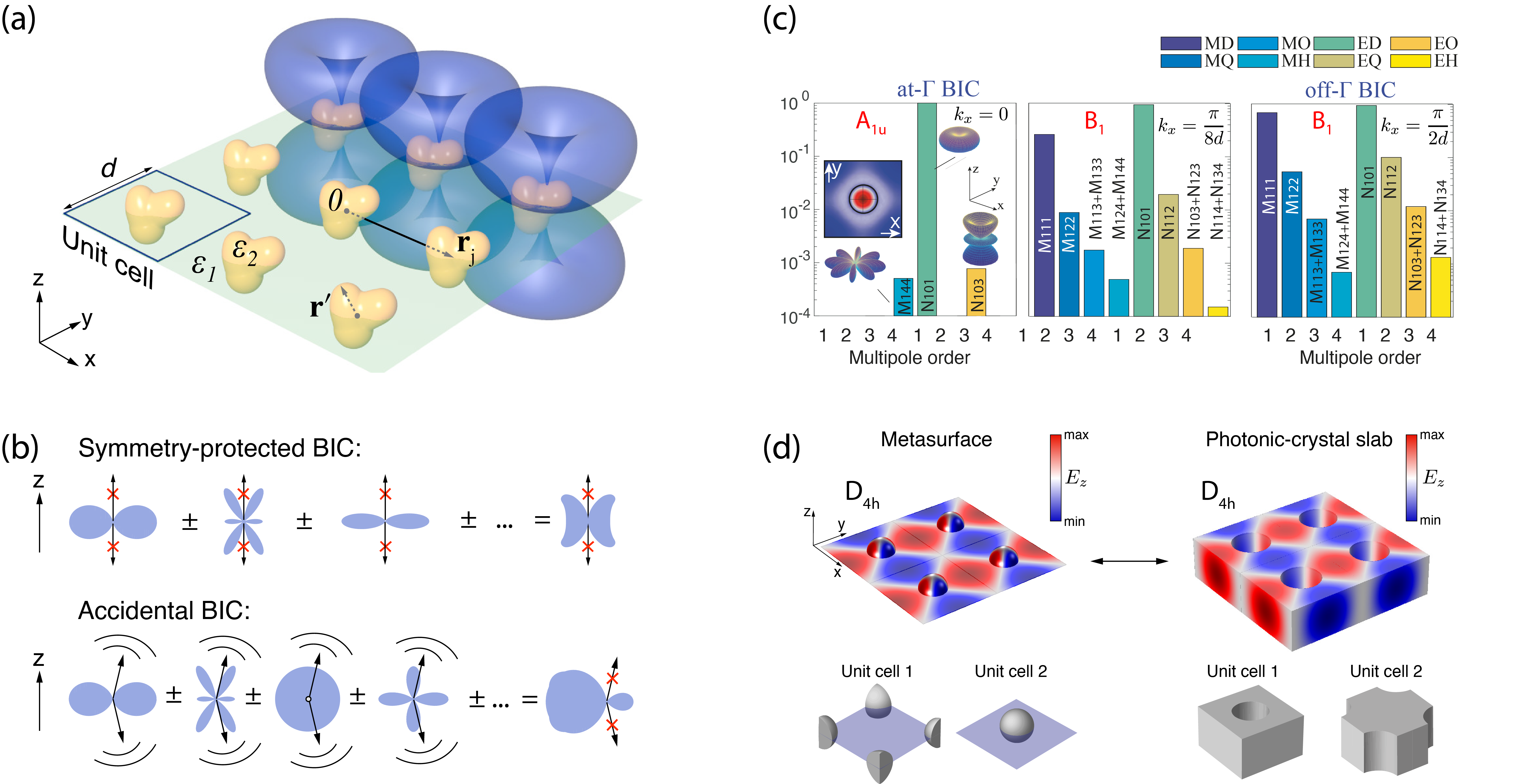}
    \caption{Multipole origin of the bound states in the continuum -- (a) A periodic dielectric metasurface with square lattice. (b) Mechanisms of BICs formation. (c) Multipole decomposition of the TM-polarized eigenmode of the metasurface composed of dielectric spheres with $\varepsilon_2 = 12$ ($\varepsilon_1 = 1$). Each panel corresponds to particular Bloch wave vector $k_x$. The vertical axis shows \textcolor{black}{the relative amplitude} 
    of electric and magnetic multipoles, \textcolor{black}{and different orders $n$ are indicated with different colors.} 
    The irreducible representations are marked by red. The inset in the first panel shows the $E_z$ field profile in the central cross section. (d) Distribution of the $E_z$ component of the electric field of the modes corresponding to the same irreducible representation $B_{1u}$ in the photonic crystal slab and metasurface. The bottom panel illustrates the variants of the unit cells for which the multipole content is identical.}
    \label{fig:2d1_v2}
\end{figure}

Here, $\{r,\theta,\varphi\}$ are spherical coordinates, $P_\ell^m$ is \textcolor{black}{an} 
associated Legendre polynomial, $z_{\ell}(k r)$ is \textcolor{black}{a} 
spherical Bessel function \textcolor{black}{describing the} 
incoming, \textcolor{black}{outgoing, } 
or standing wave, $\ell=1,2,3,...$ is the total angular momentum quantum number, $m=0,1,...\ell$ is the absolute value of the projection
of the angular momentum (magnetic quantum number), \textcolor{black}{and} $k$ is the wavenumber in vacuum or medium depending on the considered domain of the space. For \textcolor{black}{brevity}, 
we introduce $\mathbf{W}_{j}=\{\mathbf{M}_{p \ell m},\mathbf{N}_{p \ell m}\}$, where $j=\{p,\ell,m\}$. Let us consider an eigenmode of \textcolor{black}{a} photonic crystal slab \textcolor{black}{or a} 
metasurface [see Fig.~\ref{fig:2d1_v2}(a)]. The electric field of the mode outside the structure can be represented as 
\begin{equation}
    \mathbf{E}(\mathbf{r})=\sum_{\mathbf{K}_s,j}D_j  \iint\limits_{-\infty}^{\ \,+\infty}d\mathbf{k}_{\|}
    \frac{e^{i\mathbf{kr}}}{k_z}\mathbf{Y}_{j}\left(\frac{\mathbf{k}}{|\mathbf{k}|}\right)
    \delta(\mathbf{k}_b-\mathbf{K}_s-\mathbf{k}_{\|}).
    \label{eq:multipole_content}
\end{equation}
Here, $\mathbf{Y}_{j}$ is the Fourier transform of $\mathbf{W}_j$, $\mathbf{k}_b$ is the Bloch wavenumber, $\mathbf{K}_s$ is the reciprocal lattice vector, $D_j$ is the amplitude corresponding to $\mathbf{Y}_{j}$. The coefficient of \textcolor{black}{the expansion} 
of the far field $D_{j}$ is directly connected with the expansion coefficient of the polarization inside the unit cell~\cite{sadrieva2019multipolar}. 

Equation~\eqref{eq:multipole_content} \textcolor{black}{shows} 
that the far field of the structure is defined by multipole content of the unit cell, but filtered by the open diffraction channels. \textbf{The BIC is formed if the directions of all \textcolor{black}{the} open diffraction channels coincide with the nodal lines of the 
unit cell.} 

The simplest example is the subwavelength lattice of identical in-phase point dipoles oriented perpendicular to the lattice plane. Such \textcolor{black}{a} configuration correspond\textcolor{black}{s} to the mode with $\mathbf{k}_b=0$, i.e. to the $\Gamma$-point. The only allowed direction of the diffraction is normal to the metasurface, however, dipoles do 
not radiate \textcolor{black}{in} 
the direction of their axis. 
\textcolor{black}{Therefore,} this mode does not radiate, being a symmetry-protected BIC~\cite{mylnikov2020lasing}. It is shown in Ref.~\cite{sadrieva2019multipolar} that there are two scenarios of BICs in subwavelength structures [see Fig.~\ref{fig:2d1_v2}(b)]. In the first scenario, all the multipoles \textcolor{black}{contributing to}
the eigenmode do
not radiate \textcolor{black}{in} 
the direction 
orthogonal to the structure. This correspond\textcolor{black}{s} to the symmetry-protected BIC. One can show that the multipoles $\mathbf{M}_{p \ell m}$ and $\mathbf{N}_{p \ell m}$ do not radiate up- and downwards if $m\neq1$~\cite{sadrieva2019multipolar,chen2019singularities}. Therefore, all the modes in the $\Gamma$-point not containing the multipoles with $m = 1$ are symmetry-protected BICs. In the second scenario, the interference from all \textcolor{black}{the} multipoles \textcolor{black}{included in} 
the mode suppresses the radiation \textcolor{black}{in} 
the direction of \textcolor{black}{the} open diffraction channel. This correspond\textcolor{black}{s} to the accidental BIC.         



To find all \textcolor{black}{the} multipoles 
allowed for \textcolor{black}{a} certain mode, one can address to the group theory. Indeed, the eigenmodes are transformed by irreducible representations of the structure’s symmetry group~\cite{Ivchenko1995, sakoda2004optical,agranovich2013crystal}. While the unit cell determines the point group symmetry, the Bloch functions form a basis of irreducible representation of the translation group. The symmetry group of \textcolor{black}{a} particular wave vector $\vec k_b$ is defined as a subgroup of the whole point group \textcolor{black}{that} 
keeps the $\vec k_b$ invariant. Studying the multipole content of the mode, first 
we should find the irreducible representation of the mode at the given $\vec k_b$.
The set of multipoles in 
\textcolor{black}{the decomposition} is determined \textcolor{black}{directly} by \textcolor{black}{the} irreducible representation of the mode. 
Figure~\ref{fig:2d1_v2}(c) shows the multipole content of the mode supported by a square array of dielectric spheres. The parameters of the structure \textcolor{black}{are} 
described in the caption.
In the $\Gamma$ point, $\vec k_b$-group coincides with the point-group symmetry of the unit cell. Out of the $\Gamma$ point, the irreducible representation changes.  Three main multipoles \textcolor{black}{contributing to} 
the symmetry-protected BIC in the considered example are shown in Fig.~\ref{fig:2d1_v2}(c) (left panel). As we mentioned previously, for the accidental BIC, the sum of the vector spherical harmonics (Eq.~\eqref{eq:multipole_content}) is nullified in the direction of the open diffraction channel. The multipole content of the accidental BIC is shown in the third panel in Fig.~\ref{fig:2d1_v2}(c). 

One can show that in the square lattice (D$_{4h}$, D$_{4v}$) all singlet states in the $\Gamma$-point with the frequency $\omega/c<2\pi c/d$ are symmetry-protected BICs, while the bright modes are doubly degenerate
~\cite{sadrieva2019multipolar}. However, in the triangular lattice (D$_{6h}$, D$_{6v}$), there are two two-dimensional representation\textcolor{black}{s}. One of them does not 
contain the multipoles with $m=1$. Therefore, such structures can host doubly-degenerate symmetry-protected BICs~\cite{dyakov2021photonic}. Recently, Overvig and co-authors \textcolor{black}{presented} 
a quite detailed catalog\textcolor{black}{ue} of the selection rules \textcolor{black}{for} 
symmetry-protected BICs in two-dimensional photonic crystal structures~\cite{overvig2019selection}.    

If the metasurface consists of meta-atoms characterized by a single multipole (dipole, quadrupole, octupole, etc), then the position of the accidental BIC in the k-space is determined by the direction of the nodal line of the multipole. Of course, the multipole approach is more natural for the metasurfaces consisting of resonant meta-atoms, but strictly speaking, it can be applied fairly for the photonic crystal slabs 
in the case of low-contrast materials or when the filling factor is low. As an example, Fig.~\ref{fig:2d1_v2}(d) shows the field distribution for two modes of the same symmetry in \textcolor{black}{a} metasurface and \textcolor{black}{a} photonic crystal slab with D$_{4h}$ symmetry.


\subsection{BICs and topological charges}

\begin{figure}[htbp]
\centering
\includegraphics[width=0.7\linewidth]{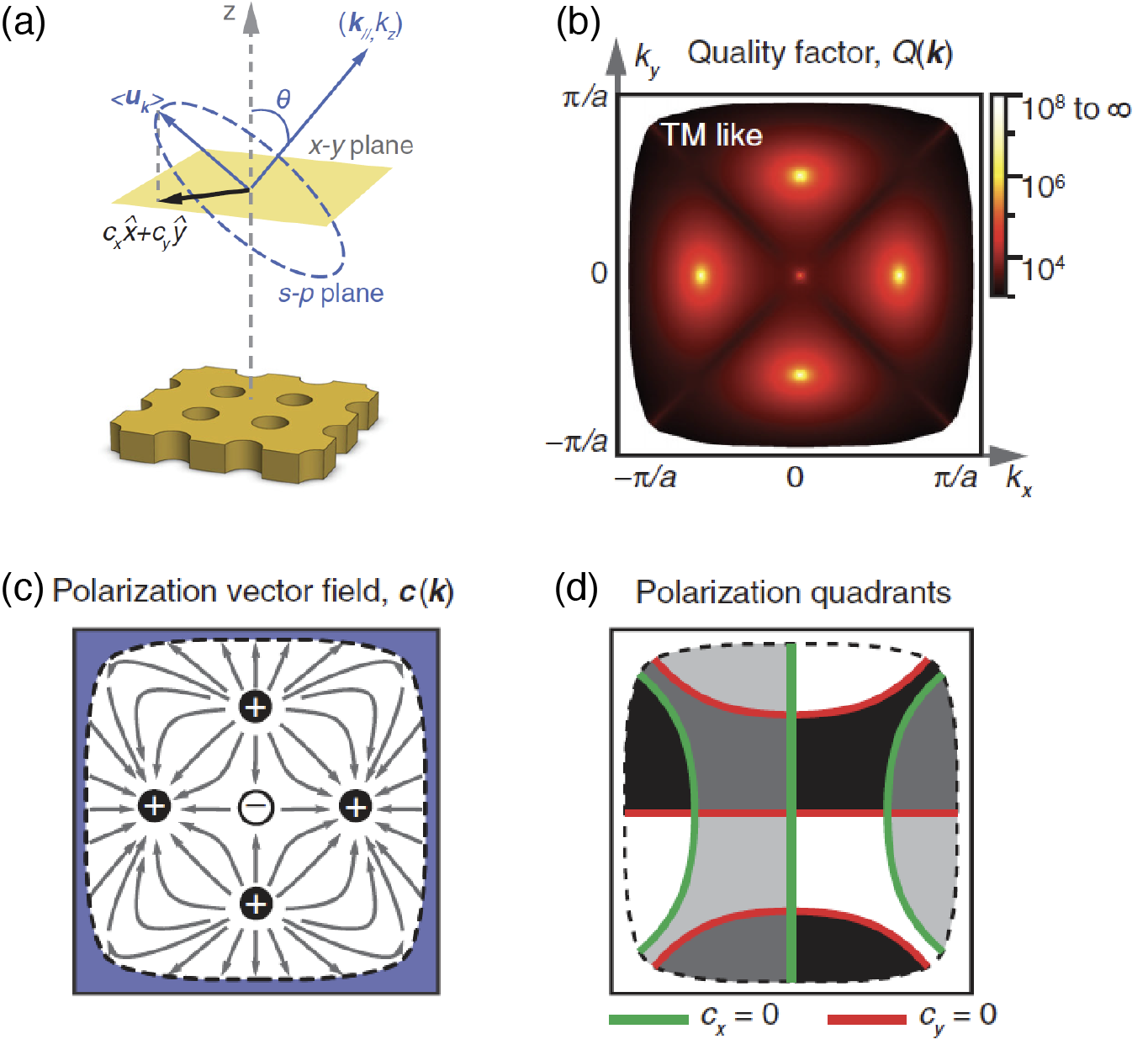}
\caption{\textcolor{black}{(a) Scheme of the expansion of the radiation field for the resonances of a photonic crystal plate. The spatially-averaged Bloch part of the electric field $\mathbf{c}(\mathbf{k})=\langle \mathbf{u}_{\mathbf{k}}\rangle$ is projected onto the $xy$ plane as the polarization vector $\mathbf {c}(\mathbf{k})=(c_x,c_y)$. Resonance turns into a BIC when $c_x=0$ and $c_y=0$. (b) Map of the radiative Q factor calculated for the TM$_1$ modes of a square-lattice photonic crystal slab in the first Brillouin zone. For the plotted mode, five BICs are visible: four accidental and one symmetry-protected. (c) Directions of the vector field of polarization show the vortices with topological charges +1 and -1. The blue shaded area indicates the region below the light line. (d) Nodal lines of the components of the polarization vector $\mathbf{c}(\mathbf{k})$. Based on the work of Zhen et al.~\cite{zhen2014topological}.}}\label{fig:2e1}
\end{figure}

Usually, BICs are robust to the change of some parameters of the system -- they \textcolor{black}{do not} 
disappear, but shift in the configuration space. Such robustness can have topological origin and, thus, can be characterized by topological invariants \textcolor{black}{(topological charges)}. Here, we consider three illustrative examples of topological robustness of BIC in the configuration space. 
\textcolor{black}{We have shown above} that BIC\textcolor{black}{s appear} in periodic structures 
when the vector Fourier amplitude responsible for the radiation to the open diffraction channels nullifies. It was shown in Ref.~\cite{zhen2014topological} that the polarization structure of the Fourier amplitude in the vicinity of \textcolor{black}{a} BIC forms a vortex that can be characterized by a topological charge showing the number of 2$\pi$ rotations of the polarization vector around the BIC \textcolor{black}{in the reciprocal space}.

\textcolor{black}{According to the Bloch theorem,} the electric field of a mode in photonic crystal slab [see Fig.~\ref{fig:2e1}(a)] can be written (similar to the case of \textcolor{black}{1D} grating considered earlier) \textcolor{black}{can be written as}
\begin{equation}
    \mathbf{E}_{\mathbf{k}}(\boldsymbol{\rho}, z)=e^{i \mathbf{k} \cdot \boldsymbol{\rho}} \mathbf{u}_{\mathbf{k}}(\boldsymbol{\rho}, z).
\end{equation}
Here, $\mathbf{k}=(k_x,k_y,0)$ is \textcolor{black}{a} 
two-dimensional two-dimensional Bloch wave vector, $\boldsymbol{\rho}=(x,y,0)$ is \textcolor{black}{a} 
radius vector lying in the plane of the slab, \textcolor{black}{and} $z$ is the normal direction to the slab.  $\mathbf{u}_{\mathbf{k}}(\boldsymbol{\rho}, z)$ is \textcolor{black}{a} 
periodic function of $\boldsymbol{\rho}$. The zero-order Fourier amplitude $\mathbf{c}(\mathbf{k})=\langle \mathbf{u}_{\mathbf{k}}\rangle$ define\textcolor{black}{s} the amplitude and polarization of the outgoing wave. The spatial average is taken over the unit cell \textcolor{black}{in} 
any horizontal plane outside the slab. Since the BIC radiates neither $x$ nor $y$ polarization, it appears in $\mathbf{k}$-space at the crossings of the line corresponding to $c_x(k_x,k_y)=0$ and $c_y(k_x,k_y)=0$ [see Figs.~\ref{fig:2e1}(b) and ~\ref{fig:2e1}(d)]. The topological charge $q$ characterizing the BIC can be introduce\textcolor{black}{d} as follows
\begin{equation}
q=\frac{1}{2 \pi} \oint_{C} d \mathbf{k} \cdot \nabla_{\mathbf{k}} \phi(\mathbf{k}), \quad q \in \mathbb{Z}.
\end{equation}
Here $\boldsymbol{\phi}(\mathbf{k})=\arg \left[c_{x}(\mathbf{k})+i c_{y}(\mathbf{k})\right]$, \textcolor{black}{and $C$ is a} 
simple path in $\mathbf{k}$-space that goes around the BIC in the counter-clockwise direction. The polarization vector has to come back to itself after the closed loop, so the overall angle change must be an integer multiple of $2\pi$, and, thus, $q$ must be an integer. For the considered structure, the symmetry-protected BIC carrie\textcolor{black}{s a charge of} 
$q=+1$, see Fig.~\ref{fig:2e1}(c). The topological charge of BIC hosted by the silicon nitride grating was measured experimentally  using the angle- and wavelength-resolved polarimetric reflectometry~\cite{doeleman2018experimental}. It was shown in Ref.~\cite{zhen2014topological} that the topological charge of BIC\textcolor{black}{s} in gratings can be only $q=0,\pm1$.  The topological charge in structures periodic in two dimensions can be arbitrary high. However, all designs of the \textcolor{black}{photonic} structures proposed  up to date support 
BICs  with \textcolor{black}{a maximum value of the topological charge of} 
$|q|=2$~\cite{yoda2020generation,zhen2014topological}. The observation of a BIC with \textcolor{black}{a} high\textcolor{black}{er} topological charge 
\textcolor{black}{remains an unsolved challenge}.

\begin{figure}[t]
\centering
\includegraphics[width=0.8\linewidth]{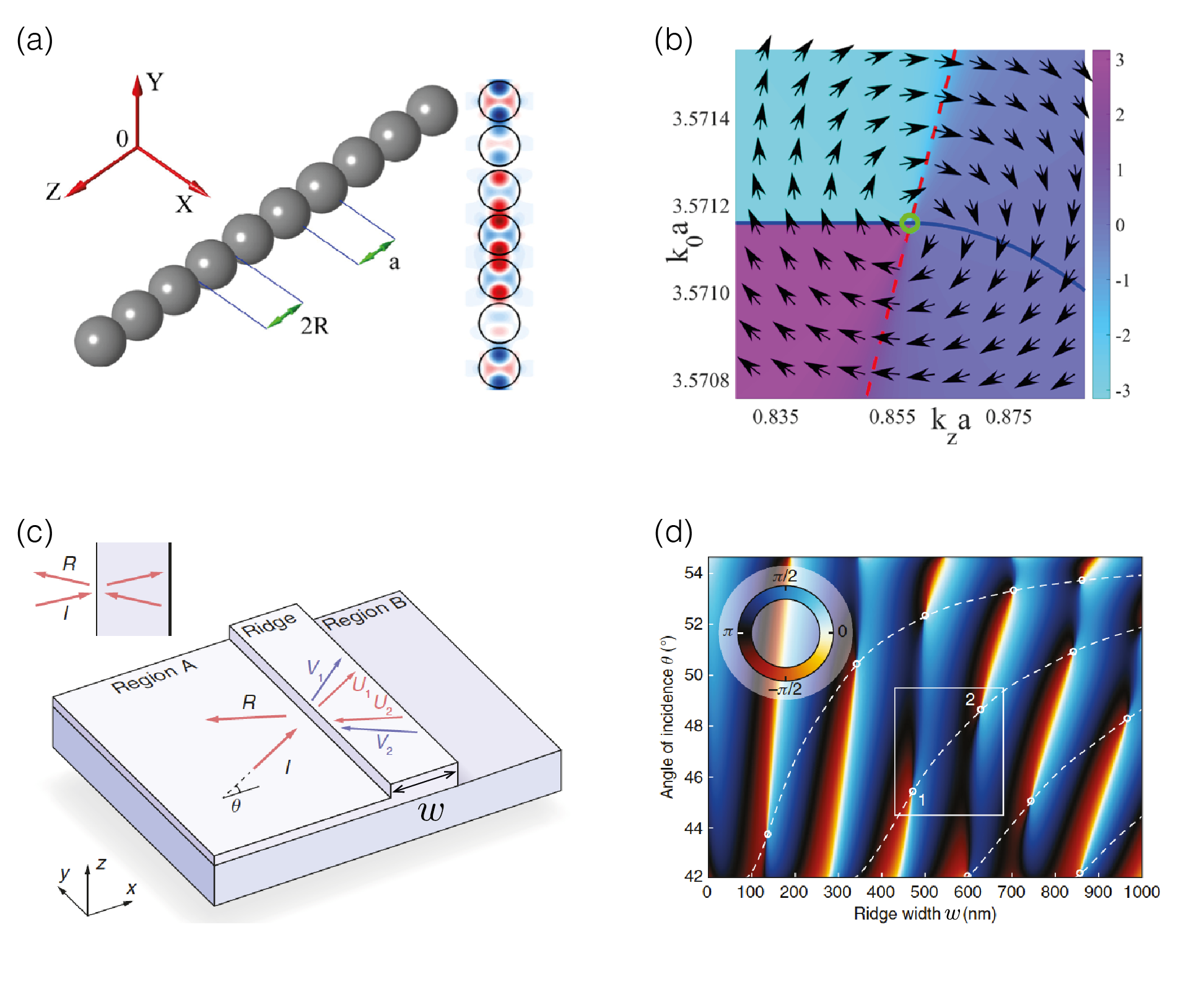}
\caption{(a)  Periodic array of dielectric spheres and \textcolor{black}{the distribution of} the magnetic field \textcolor{black}{amplitude} $H_z$  
for \textcolor{black}{a} BIC 
plotted in the $y0z$ plane. (b) Phase map of the coupling coefficient $W(\omega,k_z)$. The vectors show 
\textcolor{black}{a vortex} around a BIC. Adapted from Bulgakov et al.~\cite{bulgakov2017topological}. (c) Schematic view of the on-chip analog of the Gires-Tournois interferometer supporting a BIC. \textcolor{black}{(d)} Phase map of the coupling coefficient $\mathcal{P}(w,\theta)$. Adapted from Bykov et al.~\cite{bykov2019bound}} \label{fig:topology}
\end{figure}

Being robust against \textcolor{black}{changes in} 
some geometrical parameters of the structures, the polarization vortex can migrate inside the Brillouin zone over \textcolor{black}{a} dispersion surface. Within the given photonic branch, the total topological charge is conserved. It leads to restrictions on the behavior 
of BICs. For example, \textcolor{black}{a} BIC can be destroyed t\textcolor{black}{h}rough the annihilation when two or several topological charges with total zero charge collide. \textcolor{black}{A} BIC with an integer topological charge can decay into several BICs with integer topological charges or into circularly polarized states with 
half-integer charges~\cite{zhen2014topological,yoda2020generation,liu2019circularly,bulgakov2017bound}. Some examples of the topological charge\textcolor{black}{s'} migration and 
decay will be considered in \textcolor{black}{the} section \textcolor{black}{'Dielectric Gratings'.} 
The topological charge carr\textcolor{black}{ied} 
by \textcolor{black}{a} BIC 
can be used for the generation of optical vortex beams~\cite{huang2020ultrafast,bai2021terahertz,wang2020generating}.         

In Ref.~\cite{bulgakov2017topological}, the Authors develop\textcolor{black}{ed} 
quite \textcolor{black}{a} general approach \textcolor{black}{for defining} 
topological charge of BIC in a wide class of 
system\textcolor{black}{s} and then appl\textcolor{black}{ied} 
the developed formalism to 
linear periodic chains of coupled resonators [Fig.~\ref{fig:topology}(a)]. The Authors introduce\textcolor{black}{d} a complex function $W(\omega,k_z)$ (quasimode coupling strength) characteriz\textcolor{black}{ing} 
the coupling efficiency between the \textcolor{black}{scattering channel} 
and \textcolor{black}{the} resonant states of the system. Here, $k_z$ is the wavenumber of the incident wave and $\omega$ is its frequency. Function $W(\omega,k_z)$ can be interpreted as a projection of the incident wave onto the eigenmode, or which is the same, the projection of the resonant state \textcolor{black}{on}to the wave 
outgoing through the scattering channel. Obviously, 
$W(\omega,k_z)=0$ for BIC, \textcolor{black}{because the} 
BIC appears in $\omega-k_z$ space exactly at the crossing of line\textcolor{black}{s} corresponding to $\text{Re}\{W(\omega,k_z)\}=f(\omega,k_z)=0$ and $\text{Im}\{W(\omega,k_z)\}=g(\omega,k_z)=0$. The topological charge $q$ can be introduced as
\begin{equation}
    q=\operatorname{sgn}\left.\left(\frac{\partial f}{\partial \omega} \frac{\partial g}{\partial k_{z}}-\frac{\partial g}{\partial \omega} \frac{\partial f}{\partial k_{z}}\right)\right|_{\text{BIC}}.
\end{equation}
Figure~\ref{fig:topology}(b) shows the phase $\theta=\text{arg}\left[W(\omega,k_z)\right]$. One can see that the gradient of the phase $\mathbf{j}=\nabla \theta$ form\textcolor{black}{s} a \textcolor{black}{vortex} 
around the BIC. Therefore, \textcolor{black}{a} BIC in the linear periodic chain can \textcolor{black}{also} be characterized by a topological charge. However, it is still an open question how to measure such a charge experimentally.

Another beautiful example of BIC was observed in the dielectric ridged
waveguide forming an on-chip analog of the Gires-Tournois interferometer [see Fig.~\ref{fig:topology}(c)]~\cite{bykov2019bound}. The Authors show\textcolor{black}{ed} that such \textcolor{black}{a} 
waveguide can support a BIC -- the mode \textcolor{black}{that} 
does couple 
neither \textcolor{black}{to} waveguide modes in region A, nor \textcolor{black}{to the} propagating waves in the substrate 
\textcolor{black}{and} in air. \textcolor{black}{This type} 
of BIC was predicted and experimentally observed in \textcolor{black}{the Refs.} \cite{webster2007width,tummidi2008anomalous,nguyen2019ridge}. 
\textcolor{black}{A} rigorous theory proving that these states are BICs was developed by Bezus, Bykov, and Doskolovich~\cite{bezus2018bound}. Similar to Ref.~\cite{bulgakov2017topological}, 
\textcolor{black}{they} introduce\textcolor{black}{d} a complex quasimode coupling $\mathcal{P}(w,\theta)$ between the modes of the ridge and waveguide modes in region A. If the frequency is fixed, then the BIC in the considered system can be observed at the specific values of angles $\theta$ and width of the ridge $\omega$. Exactly at the BIC, $\left.\mathcal{P}(w,\theta)\right|_\text{BIC}=0$, and its phase \textcolor{black}{is singular (i.e. non-defined)}. 
The topological charge introduced as
\begin{equation}
q=\frac{1}{2 \pi} \oint_{\gamma} \operatorname{darg} \mathcal{P}(w, \theta)
\end{equation}
can be equal to $\pm1$. By variation of the waveguide layer thickness  in region B, the Authors show\textcolor{black}{ed that} the BICs can move in $w-\theta$ space and even annihilate, if two BICs with opposite charges collide. \textcolor{black}{We should} 
note that instead \textcolor{black}{of varying} 
$w$ at \textcolor{black}{a} fixed frequency $\omega$, one can vary $\omega$ at a fixed \textcolor{black}{ridge width} $w$. In this case, the results will be completely the same. To conclude, we 
\textcolor{black}{should add that} the topological charge can be introduced in many ways, \textcolor{black}{but} 
not all of the introduced charges \textcolor{black}{will} 
have \textcolor{black}{a} clear physical meaning, and \textcolor{black}{thus, they cannot} 
be directly measured in the experiment.

\subsection{Losses and Q factor of quasi-BICs}

\begin{figure}[t]
\centering
\includegraphics[width=0.5\linewidth]{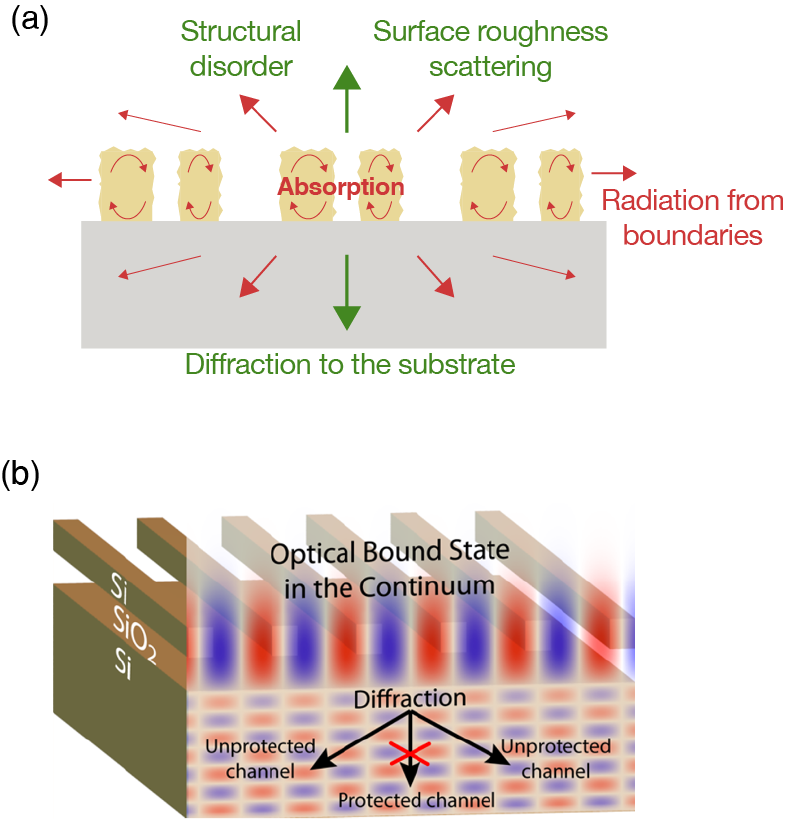}
\caption{(a) Mechanisms of losses in a periodic photonic structure. They include surface roughness, radiative loss due to structural disorder, diffraction losses due to finite size of the sample, and diffraction to high-index substrate. (b) A symmetry-protected BIC in a grating etched on the silicon-on-insulator wafer. Diffraction \textcolor{black}{that is} normal \textcolor{black}{to} 
the surface plane is prohibited by the symmetry of the BIC, while diffraction into the substrate is allowed.}
\label{fig:2g1}
\end{figure}

Losses limit the total Q-factor $Q_\text{tot}$ of BIC. \textcolor{black}{In general,} 
$Q_\text{tot}$ can be decomposed into partial contributions:
\begin{equation}
Q_\text{tot}^{-1}=\underbrace{Q_\text{rad}^{-1} + Q_\text{surf}^{-1}+Q_\text{str}^{-1}+Q_\text{size}^{-1}+Q_\text{subs}^{-1}}_{\text{radiative}}\ \ +\underbrace{Q_\text{abs}^{-1}}_{\text{non-radiative}}.
\end{equation}
Here, the radiative losses account for surface roughness ($Q_\text{surf}$), structural disorder ($Q_\text{str}$), diffraction losses due to finite size of the sample ($Q_\text{size}$), and diffraction into high-index substrate ($Q_\text{subs}$). The non-radiative losses $Q_\text{abs}$ include all type\textcolor{black}{s} of absorption (fundamental absorption, free-carrier absorption, multiphoton absorption, etc). The loss mechanisms in a periodic photonic structure are shown schemtically in Fig.~\ref{fig:2g1}(a). The statement that BICs have infinite Q factor is quite \textcolor{black}{common} 
in the literature, but it is not completely correct. Only the radiative Q factor diverges for BICs ($Q_\text{rad}\rightarrow\infty$), as they are completely decoupled from the radiative continuum, but the total Q factor can remain finite. For example, BICs with finite $Q_\text{tot}$ can be observed in  
plasmonic structures, where the \textcolor{black}{losses} 
are essential~\cite{azzam2018formation,liang2020bound,sun2021tunable}. 

Strictly speaking, BICs are \textcolor{black}{a} mathematical idealization, and unavoidable radiative losses turn them into {\it quasi-BICs}, which manifest themselves in the scattering spectra as sharp Fano resonances~\cite{koshelev2018asymmetric}. Quasi-BICs are \textcolor{black}{important for applications}, 
as they are still strongly localized and provide \textcolor{black}{a} 
giant enhancement of the incident field, \textcolor{black}{and} 
they can be excited by \textcolor{black}{an} 
external incident wave. Usually, the efficient performance of photonic structure\textcolor{black}{s} requires a critical coupling of the eigenmode to the incident field, which is achieved at the condition $Q_\text{abs}=Q_\text{rad}$~\cite{piper2014total,choi2001control,pernice2012second,seok2011radiation, koshelev2019nonlinear}. The absorption rate in semiconductors, for example, can be controlled in the visible, infrared, and terahertz ranges via free electrons induced by \textcolor{black}{an} 
external optical pulse~\cite{platte1989optically,makarov2015tuning,mazurenko2003ultrafast}. The radiative losses of quasi-BIC\textcolor{black}{s} in periodic photonic structures can be controlled by the size \textcolor{black}{of} the sample~\cite{sadrieva_experimental_2019,liu2019high,hwang2021ultralow}, \textcolor{black}{the} angle of incidence~\cite{hsu2013observation} or asymmetry of the unit cell~\cite{koshelev2018asymmetric}. It seems that the \textcolor{black}{latter is the most} 
precise and powerful \textcolor{black}{technique, and} thus, more \textcolor{black}{suitable} 
for applications requiring normal incidence excitation. The optical properties of the asymmetric photonic structures with quasi-BIC\textcolor{black}{s} and their applications are discussed \textcolor{black}{in the following Sections.} 
Another mechanism of the dynamic all-optical control \textcolor{black}{over} 
the radiative losses of BIC\textcolor{black}{s} was proposed in Ref.~\cite{chukhrov2021excitation}. It was shown that the Kerr-type nonlinearity can result 
in radiative losses, \textcolor{black}{which} appear 
due to the coupling between the symmetry-protected BIC and \textcolor{black}{a} bright mode of the system. Thus, the radiative losses of BIC\textcolor{black}{s} can be controlled by the intensity of the incident light. A similar mechanism of the radiative loss control was proposed for \textcolor{black}{the implementation of} optical memory \textcolor{black}{based} on BICs~\cite{bulgakov2015all,lannebere2015optical}.                     

One of the loss mechanisms is the diffraction into the substrate with \textcolor{black}{a} high refractive index. This mechanism is \textcolor{black}{crucial,} 
for example, for photonic structures fabricated  from \textcolor{black}{a} silicon-on-insulator wafer~\cite{sadrieva_transition_2017}. In this case, the refractive index of the substrate is higher than the effective refractive index of the BIC, thus, there is a diffraction into the substrate [see Fig.~\ref{fig:2g1}(b)]. It is important that despite the presence of the substrate with a high refractive index, the zero-order diffraction into the substrate is closed for symmetry-protected BIC\textcolor{black}{s}, as the substrate does not break $C_2$ symmetry with respect to the vertical axis. The intensity of the diffraction into the substrate strongly (exponentially) depends on the thickness of the  \textcolor{black}{SiO$_2$} buffer layer 
isolating the photonic structure from the substrate. Thus, $Q_\text{tot}$ increases with the thickness of SiO$_2$ exponentially and then saturates. The saturation plateau is defined by other loss mechanism\textcolor{black}{s}. Therefore, the higher \textcolor{black}{the} quality of the fabrication \textcolor{black}{is}, the thicker layer of SiO$_2$ is required.

The radiative loss of BICs in period\textcolor{black}{ic} structures is suppressed due to the collective destructive interference from all \textcolor{black}{the} unit cells of the structure. Thus, quasi-BIC in the experimental samples have 
radiative losses that depend 
on the number of periods $N$ (unit cells)~\cite{sadrieva_experimental_2019,bulgakov_light_2017,bulgakov_nearly_2018,bulgakov_transfer_2016,bulgakov_trapping_2017}. The asymptotic behaviour of such losses ($Q_\text{size}$) for large $N$ can be estimated 
from the known dependence of 
radiative losses in the infinite structure on the Bloch wavenumber, i.e. $Q_\text{rad}=Q_\text{rad}(\mathbf{k}_b)$. The transition from the infinite structure, \textcolor{black}{which} 
an be easily analyzed numerically or even analytically, to \textcolor{black}{a} finite one is based on the Fabry-Perot quantization of Bloch wavenumber. Thus, for the in-$\Gamma$ BIC, 
$Q_{\text{size}}$ is \textcolor{black}{approximately} 
$Q_\text{rad}(|\mathbf{k}_{b,\text{min}}|)$, \textcolor{black}{w}here $|\mathbf{k}_{b,\text{min}}|\approx\pi/L$, and $L$ is the linear size of the structure. Such \textcolor{black}{an} approximation \textcolor{black}{has proven to be useful} 
for linear chains~\cite{sadrieva_experimental_2019,sidorenko2021observation}. However, one \textcolor{black}{should} 
keep in mind that in some cases, the eigenmodes in \textcolor{black}{an} infinite lattice can substantially differ from those in finite structures even \textcolor{black}{if} their size is large~\cite{zakomirnyi2019collective}. In practice, the width of the high-Q Fano resonances ceases to depend on the sample size if it is about several hundred of periods. However, recently 
a dielectric structure \textcolor{black}{design was proposed that consists} 
of 27$\times$27 Si blocks hosting a quasi-BIC with experimentally measured $Q_\text{tot}=18500$ in the telecommunication range~\cite{liu2019high}. \textcolor{black}{However,} there is no clear recipe how to fabricate high-Q metasurfaces with a small footprint, and today it remains a \textcolor{black}{highly relevant challenge.} 
The total Q factor of BICs in periodic dielectric structures 
is about $10^3$-$10^4$ \textcolor{black}{in the visible and near-IR ranges}~\cite{dyakov2021photonic,hsu2013observation,sadrieva_experimental_2019,sidorenko2021observation,hwang2021ultralow,romano2018optical,anthur2020continuous}, and it  strongly depends on the fabrication quality. The radiative Q factor can reach values \textcolor{black}{of} $10^6$-$10^7$~\cite{hsu2013observation}. 
Recently, \textcolor{black}{a} 
symmetry-protected BIC in \textcolor{black}{a} photonic crystal slab with total Q factor of about $10^6$ was demonstrated experimentally~\cite{chen2022observation}. The radiation from the edge\textcolor{black}{s} of the structure was suppressed by surrounding the sample with a photonic crystal with another lattice constant. Thus, the frequency of BIC \textcolor{black}{was} 
in the bandgap of the surrounding photonic crystal.

Another source of 
radiative losses is the radiation induced by fabrication imperfections or structural disorder. \textcolor{black}{Such losses are} 
a common problem for high-Q photonic structures~\cite{ishizaki2009numerical,minkov2013statistics,biberman2012ultralow}. Structural fluctuations are much more difficult to control than, for example, \textcolor{black}{the} sample size. \textcolor{black}{Therefore,} 
the radiation losses due to structural fluctuation and fabrication imperfections usually are the main loss mechanism limiting the Q factor of BIC. It is well-known that the structural disorder in periodic photonic structures drastically affects their optical properties, resulting in non-trivial Fano resonance evolution, light localization, coherent back-scattering, etc ~\cite{wolf1985weak,wiersma1997localization,poddubny2012fano,limonov2012optical,liu2019disorder}. The disorder effects are most essential in self-assembled and natural photonic structures~\cite{galisteo2011self,astratov2002interplay,fan1995theoretical}.

According to the general theory, the Q factor of BIC due to the structural fluctuations behaves as $Q_\text{str}^{-1}\propto\sigma^2$, \textcolor{black}{where $\sigma$ is the disorder amplitude}~\cite{Lifshits1988}. In Ref.~\cite{ni_analytical_2017}, the Authors analyzed the radiative losses of symmetry-protected BIC\textcolor{black}{s} in 
dielectric grating\textcolor{black}{s} using the coupled-mode theory, 
finite element methods \textcolor{black}{and} 
the supercell approach. They mention that the radiation is mainly induced by fractional-order Bloch waves, particularly near the zeroth order diffraction channel. Thus, we can say that fluctuations result in BIC scattering 
to the \textcolor{black}{neighboring} states 
(in $\mathbf{k}$-space) states that are leaky. The Authors show that when the size of the unit cell becomes large enough, 
the influence of the boundary condition\textcolor{black}{s} becomes considerably weak. This can be explained by \textcolor{black}{the} spatial localization of the mode, but the Authors didn't analyze this effect. The spatial localization of the BICs \textcolor{black}{was demonstrated numerically in Ref.~\cite{Maslova2021BoundSI}, where } 
one-dimensional periodic structure composed of two layers of dielectric rods \textcolor{black}{was considered.} 
In the same work, the effect of the uncorrelated structural disorder was analyzed for both the symmetry-protected and accidental BICs. It \textcolor{black}{was} 
shown that the symmetry-protected BIC\textcolor{black}{s are} 
more resistant to the fluctuation of the rods  in the direction perpendicular to layers \textcolor{black}{compared to} 
the fluctuation of the rod position along the layers. And vice versa, the accidental BIC\textcolor{black}{s are} 
more robust against the fluctuation along the direction of periodicity.
The Anderson localiz\textcolor{black}{ation effect} 
in structures with BICs is discussed in Ref.~\cite{chen_nearly_2018}. In this work, the Authors consider\textcolor{black}{ed} the structure similar to that 
considered by Plotnik et al. [see Fig.~\ref{fig:1d1}(c)]~\cite{plotnik2011experimental} and analyze\textcolor{black}{d in detail} the stability of the BIC against disorder. They show\textcolor{black}{ed} that due to the non-trivial interplay between the BIC and disorder-induced localized states, the entangled biphoton survive\textcolor{black}{s} after averaging over all \textcolor{black}{the disorder} configurations. 

The Authors of Ref.~\cite{jin2019topologically} proposed the idea how to increase the resistance of BICs to structural disorder. They experimentally demonstrate\textcolor{black}{d} a BIC with the total Q factor about $5\times10^5$. The main idea is to merge several BICs in the k-space in the vicinity of the $\Gamma$-point. Later, merged BICs, or super-BICs, in photonic crystal slabs were used to achieve record-low threshold lasing~\cite{hwang2021ultralow}. As \textcolor{black}{we mentioned above}, 
the radiation of BIC \textcolor{black}{due to structural disorder} occurs via the fractional-order Bloch waves near the $\Gamma$-point. Thus, in the proposed design, the rescattering takes place mainly between \textcolor{black}{the} extremely high-Q states, making the BIC immune against the structural disorder.

\section{BICs in photonic structures of various dimensions}

\subsection{Gratings}


In this section, we overview the BICs in 1D dielectric gratings -- photonic structures of finite thickness periodic in one direction and having \textcolor{black}{a} 
translation\textcolor{black}{al} symmetry in other direction. The study of such structures has a long history, \textcolor{black}{which} began in 1887 from Lord Rayleigh\textcolor{black}{'s works}~\cite{rayleigh1887xvii}. The waveguide gratings were intensively studied since the middle of \textcolor{black}{the} 20-th century, and \textcolor{black}{they} found numerous applications in distributed-feedback lasers, laser mirrors, bandpass filters, wavelength demultiplexers, polarizers, chemical and bio-sensors, and other optic and optoelectronic devices. An interested reader may consult numerous reviews~\cite{elachi1976waves,suhara1986integrated,magnusson2016guided,chang2012high,quaranta2018recent}. A pronounced twist in \textcolor{black}{the} physics \textcolor{black}{of} optical gratings was related to the development of nanotechnology, namely, photonic crystals, metasurfaces, flat optics, and high-contrast gratings~\cite{qiao2018recent}.


Due to simple geometry, the BICs in dielectric gratings can be described easily using the Fourier modal method (FMM, also called rigorous coupled-wave analysis, RCWA) \cite{li2014fourier} or the true modal method (TMM)~\cite{botten1981finitely} designed to be particularly efficient for gratings, resonant state expansion~\cite{weiss2018calculate,neale2020resonant}, guided mode expansion~\cite{Andreani2006photonic}, \textcolor{black}{multiple} 
scattering theory (Korringa–Kohn–Rostoker method)~\cite{modinos2001application}, and others. FMM and TMM methods supplemented with the S-matrix technique provide a powerful tool for analysis of complex multilayer \textcolor{black}{and photonic crystal} structures \cite{cotter1995diffraction,tikhodeev2002quasiguided}. 

Along with the symmetry-protected BICs which we discussed previously, the gratings allow 
implement\textcolor{black}{ing} another type of BIC\textcolor{black}{s}, usually called {\it Fabry-Perot BIC}. It is well-known that in the vicinity of the optical resonances of non-absorbing gratings, the transmission becomes zero [see Fig.~\ref{fig:3a1}(b)]~\cite{shipman2005resonant}, i.e. the structure behaves as a perfect mirror. 
\textcolor{black}{Using} two such structures and varying the distance between them, one can completely trap the light between the gratings. 
This mechanism of BIC formation was analyzed in Ref.~\cite{marinica2008bound} [see Fig.~\ref{fig:1d1}(a)]. The Fabry-Perot BIC is a particular case of tunable (Friedrich-Wintgen) BIC, and it also can be described within coupled-mode theory by Eqs.~\eqref{eq:CMT1} and \eqref{eq:CMT2}, assuming that $\omega_1=\omega_2$ and $\gamma_1=\gamma_2$. The Fabry-Perot BIC\textcolor{black}{s} can appear at both normal $\mathbf{k}_b=0$ and oblique $\mathbf{k}_b\neq0$ incidence.

\begin{figure}[t]
\includegraphics[width=0.95\columnwidth]{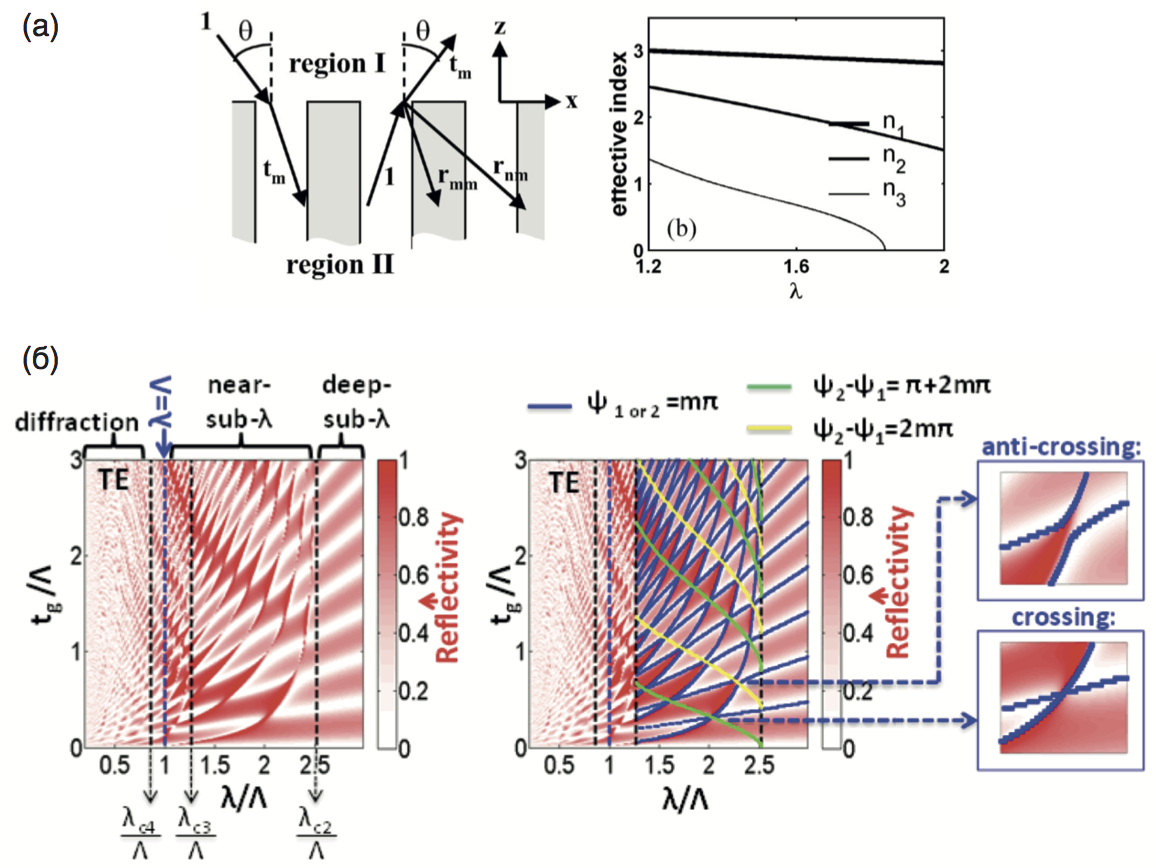}
\caption{(a) Interface between a homogeneous medium 
and a semi-infinite 1D photonic crystal (left), and spectral dependence of effective propagation constants of Bloch modes in 1D photonic crystal (right). Adapted from Lalanne et al.~\cite{Lalanne2006}. (b) Reflectivity maps for 1D photonic crystal slab: cut-off \textcolor{black}{wavelengths of} Bloch mode\textcolor{black}{s} divide the map into \textcolor{black}{areas with single-mode, two-mode, and multiple-mode regimes} 
(left); blue lines identify \textcolor{black}{the single-mode Fabry-Perot resonances},  
and green and yellow lines indicate \textcolor{black}{the} in-phase and anti-phase conditions in \textcolor{black}{the} two-mode regime (right). Adapted from Karagodsky et al. \cite{Karagodsky2012}} \label{fig:3a2}
\end{figure}

Actually, there is no need to use a double-layer structure to \textcolor{black}{obtain} a Fabry-Perot BIC. It can also appear in gratings with varying thickness. This is well-described by 
\textcolor{black}{modal methods} developed in Refs.~ \cite{tishchenko2005phenomenological,Lalanne2006,karagodsky2011matrix,karagodsky2011,Karagodsky2012}. \textcolor{black}{In these approaches,} 
Bloch modes \textcolor{black}{are used that are} 
propagating \textcolor{black}{eigen}modes of the infinite\textcolor{black}{ly} thick grating, similar to plane waves in free space~\cite{tishchenko2005phenomenological}. A plane interface between a homogeneous medium and a photonic crystal couples the modes with each other (upon reflection) and with plane waves \textcolor{black}{of the surrounding space or substrate} (upon transmission), as illustrated in Fig.~\ref{fig:3a2}(a). As the slab is bounded by two interfaces, 
multiple Fabry-Perot-like resonances appear in the reflection spectra [see Fig.~\ref{fig:3a2}(b)]. One can see that the linewidth of the resonances 
strongly depends on the thickness of the grating $t_g$, and at particular values of $t_g$ the resonances disappear from the spectra, turning into BICs. The appearance of the Fabry-Perot BICs requires the existence of two Bloch modes, at least, \textcolor{black}{in the spectrum of the infinitely thick grating}. Indeed, one can see from Fig.~\ref{fig:3a2}(b) that regardless of the slab thickness, BICs can 
exist only for $\lambda<\lambda_{c2}$. 
\textcolor{black}{This condition defines the cutoff frequency} of the second-order Bloch wave in the grating. This fact was also \textcolor{black}{used} 
by Ovcharenko et al \textcolor{black}{for the description of BICs in gratings}~\cite{ovcharenko2020bound}. They applied multimode Fabry-Perot model developed by Tishchenko~\cite{tishchenko2005phenomenological} and Lalanne~\cite{Lalanne2006} 
\textcolor{black}{and showed} that \textcolor{black}{the} multimode Fabry-Perot model accurately predicts the existence of BICs and their positions in the parameter space. Using the same formalism, Bykov \textcolor{black}{et al.} 
obtained a simple closed-form expression 
predicting the \textcolor{black}{BIC} position\textcolor{black}{s in} 
the dispersion diagrams~\cite{bykov2019coupled}. Similar results were obtained by Parriaux and Lyndin, but without any reference to BICs~\cite{parriaux2019modal}.

\textcolor{black}{A} further characterization of BICs in 1D gratings can be done in terms of S-matrix poles~\cite{weiss2011derivation,Bykov2013numerical,tikhodeev2002quasiguided,whittaker1999scattering}. The collapse of the Fano resonance for BICs means merging the zeros and poles of the S-matrix~\cite{krasnok2019anomalies}.  This condition can used to find particular parameters at which the Fabry-Perot BIC appears~\cite{bykov2019bound}.  Blanchard and co-authors \cite{blanchard2016fano} proposed a phenomenological approach based on pole-zero approximations of scattering matrix to describe the Fano resonances in the vicinity of BICs. \textcolor{black}{Combining} 
the pole and coupled-wave formalisms for 
waveguide resonances \textcolor{black}{of a grating} was done by Pietroy et al.~\cite{pietroy2007bridging}. \textcolor{black}{Authors~\cite{bonnet2003high} explained} 
the ultra-narrow resonance corresponding to the accidental BIC via strong coupling between the guided modes. 
\textcolor{black}{Using} analytically known modes of planar waveguides \textcolor{black}{as a basis}, one can quantify \textcolor{black}{describe the diversity of} 
resonances of a photonic crystal slab via the resonant state expansion method \cite{neale2020resonant}, which is also appropriate for BIC characterization \cite{neale2021accidental}.

\begin{figure}[t]
\includegraphics[width=1\columnwidth]{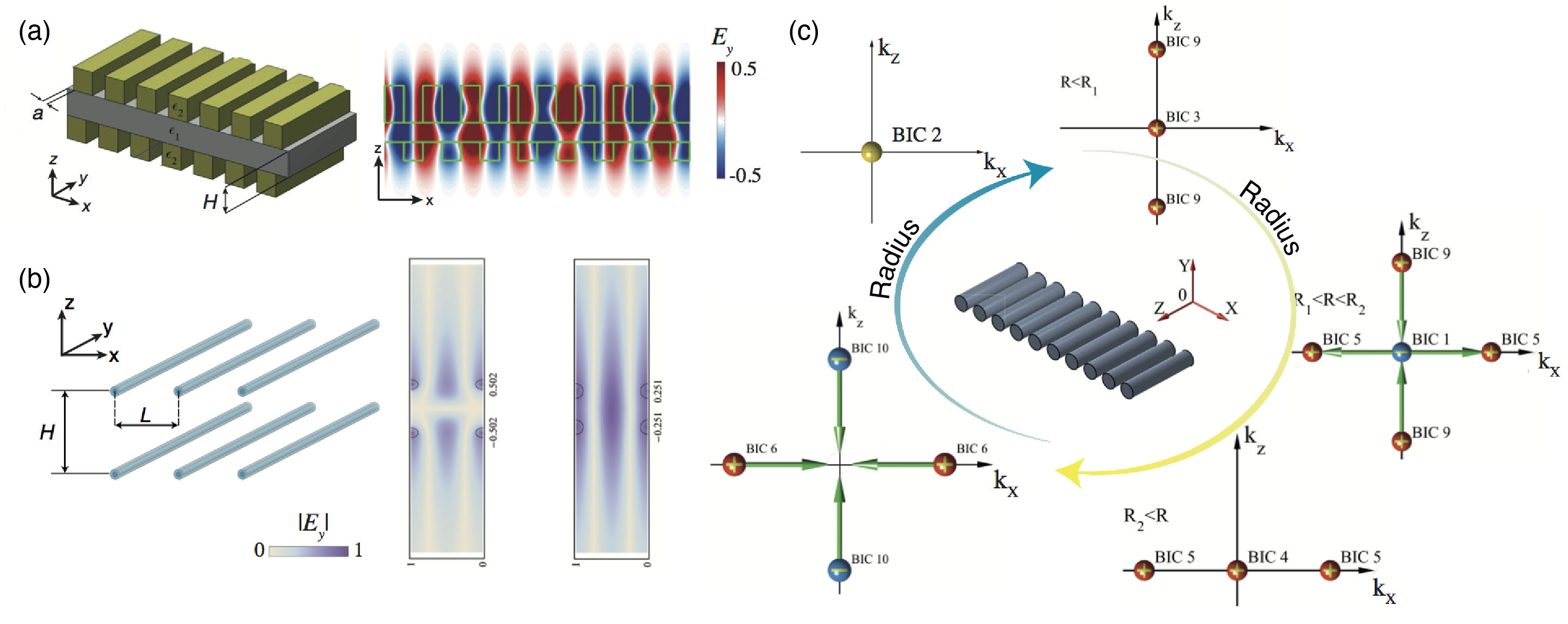}
\caption{Bound states in the continuum in one-dimensional periodic structures. (a) Schematics of a double grating and $E_y$-field profile of the accidental off-$\Gamma$ BIC. Parameter $a$ is \textcolor{black}{the} 
shift in \textcolor{black}{the} $xy$ plane, \textcolor{black}{and} $H$ is \textcolor{black}{the} 
distance between layers. Adapted from Bulgakov E.N. et al.~\cite{bulgakov2018propagating}. (b) Schematics of a periodic structure composed of two array\textcolor{black}{s} of dielectric rods, \textcolor{black}{and the} $E_y$-field profile of the accidental in-$\Gamma$ BICs. Adapted from Ndangali 
et al.~\cite{ndangali2010electromagnetic}.(c) Topological charges migration in a grating composed of dielectric rods. Yellow and blue arrows depict two different eigenmodes. Adapted from Bulgakov E.N. et al.~\cite{bulgakov2017boundsing}} \label{fig:3a3}
\end{figure}

Previously, we consider\textcolor{black}{ed} the gratings with up-down mirror symmetry. However, BIC\textcolor{black}{s} can exist in 
grating\textcolor{black}{s} without such a symmetry. \textcolor{black}{This} 
problem was considered by Ndangali and Shabanov~\cite{ndangali2010electromagnetic} and Bulgakov et al.~\cite{bulgakov2018propagating}. Bulgakov and coauthors consider\textcolor{black}{ed a} 
grating consist\textcolor{black}{ing} 
of a slab with ridges periodically arranged either on top or on both sides of the slab [see Ref.~\ref{fig:3a3}(a)]. In the case of two gratings, \textcolor{black}{they are assumed to} 
have the same period, but \textcolor{black}{different permittivities, and they} can be shifted with respect to each other by distance $a$. The \textcolor{black}{Authors} demonstrated that if a two-sided grating possesses either mirror symmetry with  respect to the $xy$-plane or glide symmetry, i.e. a composition of a mirror reflection in the $xy$-plane and a half-period translation along the $x$ axis, \textcolor{black}{then} the BIC are stable \textcolor{black}{against} 
variation of parameters as long as these 
symmetries are preserved. \textcolor{black}{In this case,} only a shift of the BIC along the dispersion curve occurs. If the up-down symmetry is broken due to the different geometries  \textcolor{black}{or material parameters} of the lattices, the existence of the accidental off-$\Gamma$ BIC requires  \textcolor{black}{a} very fine adjustment of the system's parameters. Thus, such BICs are not robust against variation of the material or geometrical parameters of the structure. The obtained results are in complete accordance  \textcolor{black}{with those} 
obtained earlier for by Ndangali and Shabanov, who consider\textcolor{black}{ed} double arrays of thin dielectric rods shifted with respect to each other~\cite{ndangali2010electromagnetic} shown in Fig.~\ref{fig:3a3}(b). They conclude\textcolor{black}{d} that in the general case of an arbitrary shift  \textcolor{black}{of one lattice with respect to the other one results in} 
the symmetry-protected BIC turn\textcolor{black}{ing} into the accidental BIC, and the accidental BIC 
turn\textcolor{black}{ing} into  \textcolor{black}{a} high-Q quasi-BIC. In addition, the Authors show that BICs can exist, if two or three diffraction channels are open, but the existence of the\textcolor{black}{se} BICs requires  \textcolor{black}{an} adjustment of  \textcolor{black}{the} radius and  \textcolor{black}{permittivity} 
of the rods.

\textcolor{black}{Ndangali and} Shabanov
~\cite{ndangali2010electromagnetic} consider\textcolor{black}{ed} the BICs in \textcolor{black}{an} 
array of very thin rods, \textcolor{black}{where each rod} 
can be described within the Rayleigh approximation, i.e. $\sqrt{\varepsilon} k_0 R\ll1$, where $k_0$ is the wavevector in vacuum,  $R$ is the radius of the rods, and $\varepsilon$ is their permittivity. Bulgakov and Sadreev generalized the considered problem to the case of arbitrary radius $R$ and permittivity $\varepsilon$, using the multiple scattering theory in the T-matrix formalism~\cite{yasumoto2018electromagnetic,linton1996scattering,twersky1952multiple}. They consider\textcolor{black}{ed} an array of GaAs rods with $\varepsilon=12$. The size parameter $\sqrt{\varepsilon}k_0R$ was in the range \textcolor{black}{from 1 to 10.} 
They show\textcolor{black}{ed} that 
such a system supports three types of BIC: (i) symmetry-protected \textcolor{black}{BICs} with a zero Bloch vector, (ii) \textcolor{black}{BICs} embedded in one diffraction channel with nonzero Bloch vector, (iii) and \textcolor{black}{BICs} embedded in two and three diffraction channels. The first and second \textcolor{black}{BIC} types 
exist for a wide range of material parameters of the rods, while the third \textcolor{black}{one} occurs only at a specific value of 
the rods' \textcolor{black}{radius or permittivity}. Yuan and Lu considered a similar system numerically and \textcolor{black}{determined} 
the domains of the existence of BICs in the parametric \textcolor{black}{space} 
$(R,\varepsilon)$~\cite{yuan2017propagating}. Bulgakov and Maksimov stud\textcolor{black}{ied} 
finite arrays consist\textcolor{black}{ing} 
of $N$ parallel dielectric \textcolor{black}{rods} and analyzed the dependence of quasi-BICs and \textcolor{black}{resonances below the} continuum  (guided-mode resonances) on $N$~\cite{bulgakov_light_2017}. They identified two types of BICs with radiative Q factors scaled as $Q\sim N^2$ and $Q\sim N^3$.

As we discussed previously, 
BICs in periodic structure\textcolor{black}{s} can be considered as a polarization v\textcolor{black}{o}rtex with a certain topological charge, and this vortex can demonstrate non-trivial dynamics in $\mathbf{k}$-space with the variation of the system's parameters, including annihilation, merging, and decay. All \textcolor{black}{these} 
effects were demonstrated \textcolor{black}{in the Ref.~\cite{bulgakov2017boundsing}  using} 
the example of BICs in \textcolor{black}{a} 
periodic array of rods. Figure~\ref{fig:3a3}(c) shows the evolution of topological charges 
for two families of BICs~\cite{bulgakov2017boundsing}. 
\textcolor{black}{Following} the yellow arrow corresponding to the increase of the radius of the rods, one can see that the symmetry-protected (BIC 3) with $q=+1$ decays into two accidental BICs (BIC 5) with $q=+1$ and one symmetry-protected BIC (BIC 1) with  $q=-1$. Further increase of the radius results in merging of two accidental BICs (BIC 9) in the $\Gamma$-point with the symmetry-protected BIC (BIC 1).  It \textcolor{black}{is} worth mentioning that due to $\omega(\mathbf{-k})=\omega(\mathbf{k})$ symmetry, off-$\Gamma$ BICs emerge and disappear in pairs. Annihilation of topological charges in the $\Gamma$ point may lead to \textcolor{black}{the} creati\textcolor{black}{on} 
of both accidental and symmetry-protected BICs. \textcolor{black}{The type of the forming BIC is defined} 
by the total topological charge conservation law. Figure~\ref{fig:3a3}(c) (evolution along the blue arrow) illustrates the creation of the accidental in-$\Gamma$ BIC $q=0$  when two BICs labelled as BIC 6 with $q=+1$ 
merge with two BICs labelled as BIC 10 with $q=-1$.

The value of topological charge determines \textcolor{black}{the} asymptotic dependence \textcolor{black}{of} the radiative  Q factor on the Bloch wavenumber. Due to the $C_2$ symmetry, one-dimensional periodic structures possess BICs with topological charges \textcolor{black}{$q=0$ or $q= \pm 1$}.
~\cite{zhen2014topological}.
The \textcolor{black}{Q} 
factor of \textcolor{black}{an} isolated BIC with $|q| = 1$ scales as $Q\sim 1/(k\pm k_\text{BIC})^2$~\cite{yuan2017strong, bulgakov2017boundsing,jin2019topologically}. For example, the Q factor of the symmetry-protected BIC \textcolor{black}{decreases as $k^{-2}$}. 
Approaching the point of the annihilation, the dependence changes to $Q\sim 1/(k-k_\text{BIC})^2 (k+ k_\text{BIC})^2$ for both accidental in-$\Gamma$ and symmetry-protected BICs. The inverse fourth-power relation indicates that the quality factor can be very large, even when $|k-k_{BIC}|$ is not 
small~\cite{yuan2017strong}. However, if the system parameters are detuned from \textcolor{black}{this} 
regime, the \textcolor{black}{Q factor of the} symmetry-protected BICs 
\textcolor{black}{decreases as} $Q\sim 1/k^2$ \textcolor{black}{again. These} 
asymptotic dependencies are extremely important, as they determine the radiative losses of quasi-BICs in finite-size structures and their robustness to disorder~\cite{jin2019topologically,sadrieva_experimental_2019,sidorenko2021observation}.

 
 

\subsection{1D periodicity with axial symmetry}

\begin{figure}[t]
\includegraphics[width=1\linewidth]{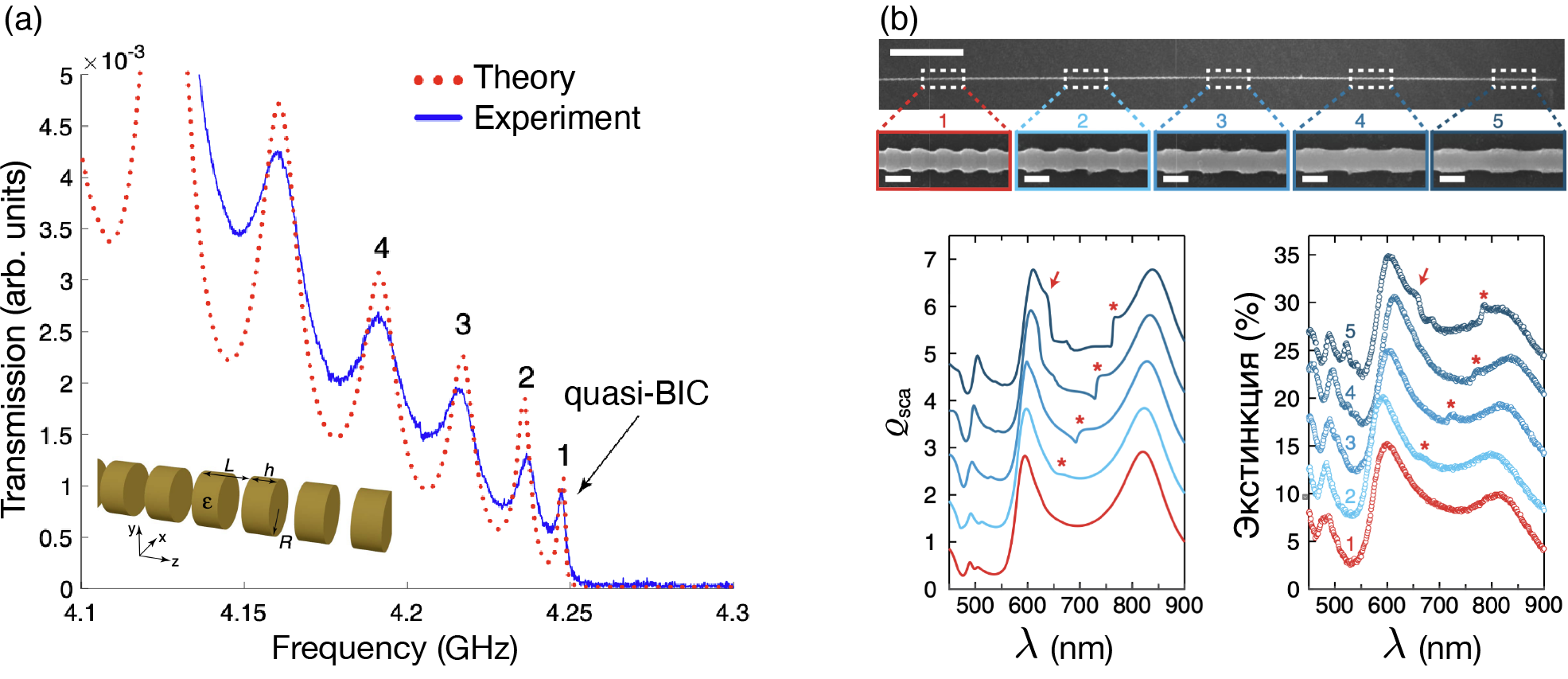}
\caption{(a) Experimental transmission spectra of \textcolor{black}{a} 
chain consisting of 20 ceramic 
disks placed between two coaxially positioned loop antennas shown in the inset. The dotted line shows the results of numerical simulations carried out in COMSOL MULTIPHYSICS. The last peak in the series corresponds to quasi-BIC. Adapted from Sadrieva et al.~\cite{sadrieva_experimental_2019}. (b) Upper: SEM image of a nanowire containing five superlattices
sections; \textcolor{black}{the} scale bar \textcolor{black}{is} 10 $\mu$m. Bottom: simulated scattering efficiency spectra (left) and experimental extinction spectra (right) (spectra offset by 5$\%$) of \textcolor{black}{the} geometric superlattices. Adapted from Kim et al.~\cite{kim2019optical}. } \label{fig:3b1}
\end{figure}


1D periodic structures with 
axial symmetry [corrugated cylindrical waveguides, \textcolor{black}{and} chains of spheres or disks (see the inset in Fig.~\ref{fig:3b1}(a)]  can also support different types of BICs. The theory of BICs in such structures was formulated by Bulgakov and Sadreev~\cite{bulgakov_trapping_2017} and further developed in Refs.~\cite{bulgakov2017bound,bulgakov2017topological,gao2019bound}. Owing to the axial symmetry and periodicity of the structure along the $z$-direction, the solution can be written in the following form
\begin{equation}
    \textbf{E}(r,\varphi,z,t) = \textbf{u}_{m,k_z}(r, z)e^{-i\omega t\pm ik_z z\pm im\varphi},
\end{equation}
where $\varphi$ is an azimuthal angle, $m$ is the azimuthal quantum number, $k_z$ is the Bloch wavenumber, and $\textbf{u}_{m,k_z}$ is \textcolor{black}{a} 
periodic function of $z$. In the special case of $m=0$, the solutions of Maxwell's equation\textcolor{black}{s} in cylindrical coordinates can always \textcolor{black}{be} divided into $TE$ and $TM$ polarizations~\cite{snyder2012optical}. It \textcolor{black}{is} worth mentioning that in contrast to uncorrugated waveguides, in our case, the solutions do not split into two independent polarizations for $k_z=0$, 
as $k_z$ is the quasi-wavenumber which is  determined up to the reciprocal lattice vector. Thus, for \textcolor{black}{$k_z=0$ and} $m\neq0$, all the modes have hybrid TE-TM polarization, as well as for $m=0$ and $k_z\neq0$.
Therefore, the modes with $k_z=0$ and $m=0$ have only one open diffraction channel in subwavelength structures. \textcolor{black}{For the odd modes with respect to the reflection in the $xy$ plane, see Fig.~\ref{fig:3b1}(a),} the coupling to this diffraction channel can vanish \textcolor{black}{because of} 
symmetry reasons, 
\textcolor{black}{similar to the case of} dielectric gratings. Therefore, 
subwavelength structures with \textcolor{black}{a} symmetric potential ($\varepsilon(-z)=\varepsilon(z)$) support the symmetry-protected BICs in the $\Gamma$-point. For the modes with $m\neq 0$ at the $\Gamma$-point, the radiative losses to 
one channel (TE or TM) can vanish \textcolor{black}{due to the symmetry of the mode.} 
The losses to the second channel can be nullified by fine adjustment of the system's parameters. Bulgakov and Sadreev named 
such states partially symmetry-protected BICs in the work~\cite{bulgakov2017trapping}. 
They \textcolor{black}{also} demonstrate\textcolor{black}{d} that BICs with $m\neq0$ and $k_z\neq0$ can be created by fine tuning \textcolor{black}{of} the chain parameters. As we mentioned previously, BICs in the periodic chain can be characterized by the topological charge, \textcolor{black}{similar to} 
BICs in 2D periodic structures. However, it is still unclear how to measure this charge experimentally. 

BIC\textcolor{black}{s} in \textcolor{black}{a} 
periodic chain \textcolor{black}{were} observed experimentally for the first time in \textcolor{black}{the} GHz frequency range, 
\textcolor{black}{in an} 1D array of coupled ceramic disks [see Fig.~\ref{fig:3b1}(a)]~\cite{sadrieva_experimental_2019}. To observe the symmetry-protected BIC, two identical loop antennas \textcolor{black}{were used as magnetic dipoles: they were} placed coaxially with the chain and connected to ports of a vector network analyzer. The measured and calculated transmission spectra are shown in Fig.~\ref{fig:3b1}(a). The last peak in the series correspond\textcolor{black}{s to a} 
quasi-BIC, which turns into a genuine 
BIC in the infinite chain. It was shown that the total Q factor of the quasi-BIC grows quadratically with the increase of the number of periods and then saturates at the level $Q=4000$ due to the material absorption. \textcolor{black}{An} 
accidental BIC ($m=0$, $k_b=0$)  was observed experimentally in a similar system in Ref.~\cite{sidorenko2021observation}. The linear growth of Q factor with the number of disk\textcolor{black}{s was} 
demonstrated for the observed accidental off-$\Gamma$ BIC. On the contrary, the radiative $Q$ factor of the accidental in-$\Gamma$ BIC follows the scaling law as $Q \sim N^3$ with $N$ being \textcolor{black}{the} 
number of scatterers in the array~\cite{bulgakov_light_2017}.

The first experimental observation of a BIC in \textcolor{black}{an} 
axially symmetric periodic structure in the optical \textcolor{black}{range was} 
presented in 2019~~\cite{kim2019optical}. The Authors fabricate\textcolor{black}{d a} silicon nanowire with \textcolor{black}{a} periodic grating at the surface as shown in Fig.~\ref{fig:3b1}(b). The surface grating forms a \textcolor{black}{structure} 
similar to a periodic \textcolor{black}{array} 
of disks placed coaxially on a core.
Since the nanowire \textcolor{black}{was} 
illuminated by a plane wave at normal incidence, 
quasi-BICs 
with different $m$ \textcolor{black}{were} 
excited. \textcolor{black}{The Authors analyzed} both theoretically and experimentally 
how \textcolor{black}{the} appearance of a BIC depends 
on illumination and 
geometry of cylindrical nanowire. The BIC-induced light confinement \textcolor{black}{in such structures} can be used to substantially enhance absorption, which can be \textcolor{black}{applied} 
in photodetector or photovoltaic devices incorporating a p–i–n diode design \textcolor{black}{and} based on nanowires with superlattice\textcolor{black}{s}~\cite{Cahoon2019}.

Nanowires with \textcolor{black}{a} periodic superlattice, and array\textcolor{black}{s} of disks have advantages over the array of spheres \textcolor{black}{due to a larger number of degrees of freedom}. By independent adjustment of the period, height, and radius of the disks, one can provide 
precise mode engineering, 
\textcolor{black}{and obtain several} BICs with different orbital angular momenta and Bloch vectors. Such linear chains supporting BICs can be used as a compact source of optical beams with angular momentum.

\subsection{2D periodicity. Photonic crystal slabs}

In this section, we review the main properties of BICs in photonic crystal slabs (\textcolor{black}{waveguides}) - 2D dielectric photonic structures that have a band gap for \textcolor{black}{the waves propagating in the waveguide plane} 
and that use index guiding to confine light in the third dimension. 
Since 1990s, such structures are considered 
\textcolor{black}{easier} for fabrication than photonic crystals with full three-dimensional band gaps, while retaining many of the latter’s desirable properties~\cite{meade1994novel,johnson1999guided}. Compared to 1D structures, photonic crystal slabs provide a broader variety of designs for unit cells, different types of their arrangement and, consequently, \textcolor{black}{more degrees of freedom for flexible control over their} 
optical properties~\cite{fan2002analysis,Andreani2006photonic}. Due to this fact, \textcolor{black}{the variety of} 
BICs in such structures \textcolor{black}{is} 
more extensive than for one-dimensional gratings or chains, including \textcolor{black}{the} possibility of \textcolor{black}{BICs with} 
high-order ($\ge 2$) topological charges~\cite{zhen2014topological} \textcolor{black}{and} robustness to disorder~\cite{jin2019topologically}.

\begin{figure}[t]
\includegraphics[width=0.9\columnwidth]{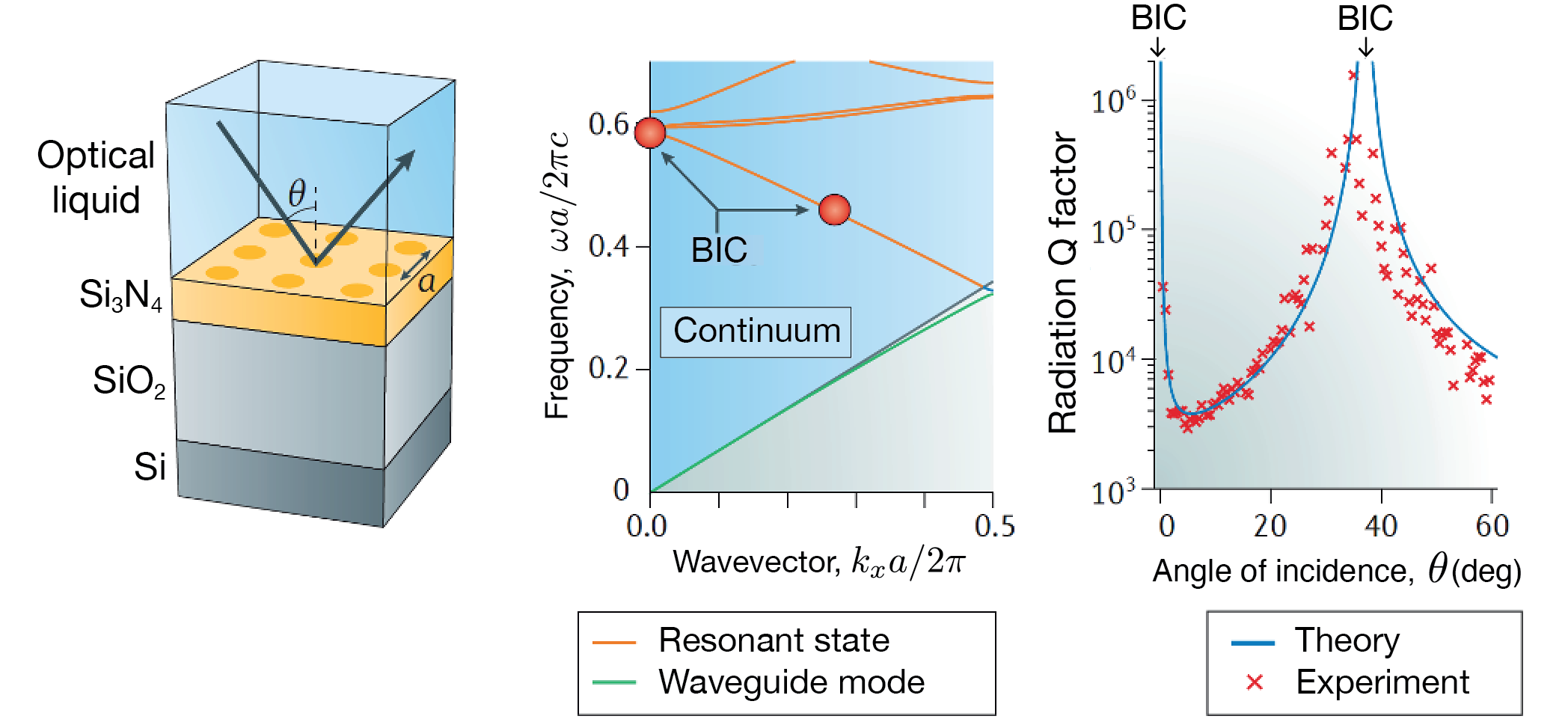}
\caption{Left: schematic layout of the fabricated structure, 
immersed in a liquid, index-matched to silica at $740$ nm. Center: The band structure along $\Gamma$-X direction with TM polarization. The BICs are marked with red dots. Right: normalized radiative lifetime extracted from the experimentally measured reflectivity spectrum (red \textcolor{black}{crosses}). The blue solid line shows the prediction from FDTD. Adapted from Hsu et al.~\cite{hsu2013observation}}
\label{fig:3c2}
\end{figure}

The first systematic study of BICs in photonic crystal slabs was performed in 2013~\cite{hsu2013observation} for a Si$_3$N$_4$ slab patterned with a periodic arrangement of circular holes as shown in Fig.~\ref{fig:3c2}, left panel. The slab was grown on a SiO$_2$/Si substrate and immersed in a liquid, index-matched to silica, which was important to keep the up-down reflection symmetry required for observation of accidental BICs. The band structure for TM-like modes along the $\Gamma$-X direction is shown in Fig.~\ref{fig:3c2}, central panel. 
\textcolor{black}{The structure supports a symmetry-protected BIC} at the normal incidence and \textcolor{black}{an accidental BIC at a non-zero Bloch vector corresponding to an} oblique angle of incidence around $35$ degrees, as can be seen in Fig.~\ref{fig:3c2}, right panel. The red crosses show the values of the radiative Q factor calculated from the total Q factor, \textcolor{black}{which was} extracted from the measured reflectiv\textcolor{black}{ity} 
spectra by assuming the value of non-radiative Q factor equal to $10^4$. The achieved value of radiative Q factor for the accidental BIC reaches $10^6$; however, the value of \textcolor{black}{the} total Q factor remains limited to $10^4$ because of parasitic and non-radiative losses. This work was the first direct evidence of true accidental BICs measured experimentally, to the best of our knowledge.

 \textcolor{black}{The} 
  \textcolor{black}{physics and origins of accidental BICs were analytically described} \textcolor{black}{in Ref.~\cite{yang2014analytical} within a coupled-wave theory (CWT) developed specially for description of response for 2D photonic crystal slabs 
  ~\cite{liang2011three}. Within this approach, the \textcolor{black}{electromagnetic fields of} eigenmodes of \textcolor{black}{a photonic system} 
  are expanded into guided modes 
  and radiative modes 
  }\textcolor{black}{of a uniform waveguide, which are used as basis functions. Due to the interaction with the lattice, the guided modes also have radiative losses. Thus, both the guided and leaky modes can contribute to the same radiation channels (diffraction channels) due to the interaction with the periodic potential of the structure. In the work~\cite{yang2014analytical}, it was shown that accidental BICs are formed due to the destructive interference of all the basis functions: both the guided and leaky ones. Two reasons of destructive interference were indicated. First, the contributions from equivalent directions in the reciprocal space can result in spontaneous appearance of an additional symmetry and, thus, to the formation of an accidental BIC. An example of such a BIC is presented in the work~\cite{hsu2013observation}. Second, it was shown that accidental BICs do not necessarily occur exactly at a high-symmetry point in the reciprocal space. The reason is, in contrast to symmetry-protected BICs, for accidental BICs, the guided basis modes contribute to the open radiation channels of the photonic structure with different weights and, therefore, after the radiation is cancelled at the symmetry point, some residual radiation can still remain. In other words, the point corresponding to a BIC in the reciprocal space does not always have a high symmetry, but is located near such a point. This is typical for the Friedrich-Wintgen mechanism~\cite{friedrich1985physical,rybin2017high}.} 
 Furthermore, it was shown analytically \textcolor{black}{that} the locations of accidental BICs can be shifted by changing various parameters, such as the cladding permittivity or geometrical sizes \textcolor{black}{of the waveguide}. It is worth mentioning that \textcolor{black}{in other works,} 
 analytical considerations demonstrate that the energy of symmetry-protected BICs is confined dominantly in the \textcolor{black}{closed} diffraction channel\textcolor{black}{s of the $\pm$ orders.} 

The moderate value of the total Q factor of $10^4$ in Ref.~\cite{hsu2013observation} is related to the out-of-plane losses originating in realistic samples due to finite size of samples and fabrication imperfections. Recently, a new approach was suggested, \textcolor{black}{which allows suppressing}  
the out-of-plane losses, \textcolor{black}{using} 
the topological nature of BICs, \textcolor{black}{namely,} by merging several BICs at the $\Gamma$ point in the k-space~\cite{jin2019topologically,koshelev2019light}. In the paper, the Authors considered a \textcolor{black}{2D} dielectric 
\textcolor{black}{photonic crystal}  
membrane suspended in air. The fundamental TE-band of the membrane supported one symmetry-protected BIC at the $\Gamma$ point and eight accidental BICs located symmetrically around the \textcolor{black}{$\Gamma$ point}. 
The tunability of accidental BICs allows \textcolor{black}{moving} 
them \textcolor{black}{away from an} off-$\Gamma$ \textcolor{black}{position} 
towards \textcolor{black}{the center of the Brillouin zone}. 
The experimentally measured value of $5\times10^5$ was demonstrated for the total Q factor in the merging regime. \textcolor{black}{For a single isolated BIC with a $\pm 1$ charge,} the Q factor of standard symmetry-protected BICs 
decay\textcolor{black}{s} quadratically ($Q \sim 1/k^2$) with respect to the distance $k$ \textcolor{black}{from the $\Gamma$ point}. 
However, it was shown that the scaling changes to $Q \sim 1/k^6$ in the configuration in which all nine BICs merge. Later, it was shown that for finite-size samples, the highest Q factor can be achieved not for the complete merging condition, but in the pre-merging regime, \textcolor{black}{when} 
the accidental BICs lie on the circle of \textcolor{black}{a} small radius in the k-space~\cite{hwang2021ultralow}.

One of the remarkable properties of 2D periodic photonic structures is that a BIC can exist even in chiral samples without the in-plane inversion symmetry~\cite{han2021extended}. In general, a BIC in the sub-diffractive regime can be achieved via cancellation of radiation into both polarization channels. In Ref.~\cite{han2021extended}, the Authors showed that by using a cross-shape\textcolor{black}{d} periodic photonic structure with\textcolor{black}{out a second-order symmetry axis}, 
it is possible to cancel radiation \textcolor{black}{amplitudes} for both polarizations: one polarization channel is suppressed  \textcolor{black}{due to the vertical mirror plane}, 
and another one parametrically, 
tuning \textcolor{black}{the} geometry of the unit cell. \textcolor{black}{This mechanism of BIC formation is very similar to the one considered by Bulgakov and Sadreev for obtaining partially symmetry-protected BICs in a chain of dielectric spheres\cite{bulgakov2017trapping}.}



\subsection{Individual subwavelength  \textcolor{black}{resonators}}

\begin{figure}[t]
\includegraphics[width=0.72\linewidth]{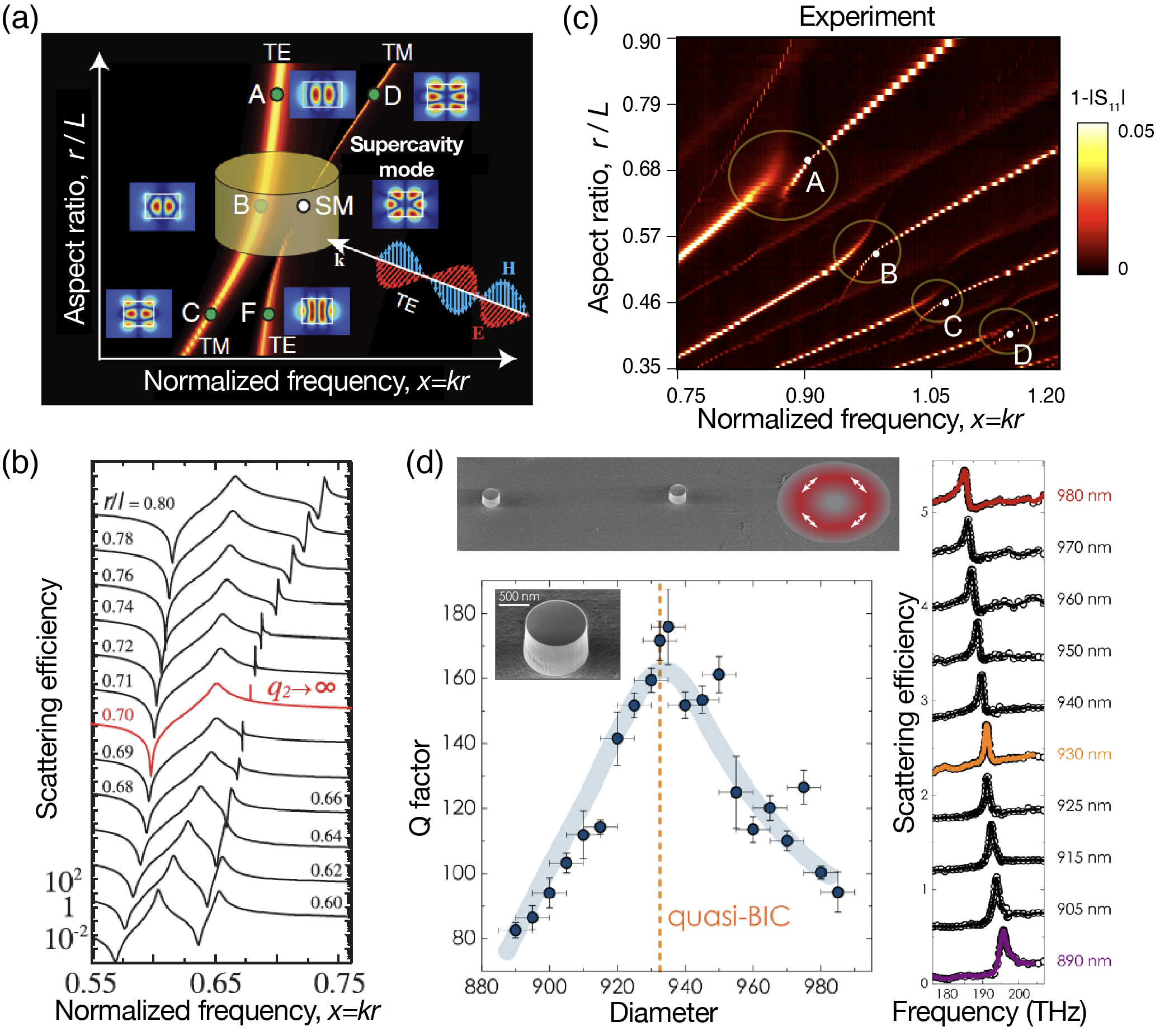}
\caption{(a) Illustration of strong mode coupling and a bound state in the continuum supported by a high-index dielectric resonator. (b) Simulated scattering spectra of a \textcolor{black}{nanodisk with $\varepsilon=80$ and azimuthal quantum number $m=0$.} 
Adapted from Rybin et al.~\cite{rybin2017high}. (c) Measured map of the coefficient $1 - |S_{11}|$ versus frequency $kr$ and aspect ratio $r/L$ for a ceramic nano\textcolor{black}{disk} 
with $\varepsilon=44.8$. The \textcolor{black}{quasi-BICs} 
labeled as A–D correspond to aspect ratios $r/L = 0.71$, $0.55$, $0.47$, and $0.42$, \textcolor{black}{respectively}. Adapted from Odit et al.~\cite{odit2021observation}. (d) Upper: SEM image of several isolated nano\textcolor{black}{discs and electric field distribution for an azimuthally polarized incident wave.} 
Lower: \textcolor{black}{dependence of the experimental-}extracted Q factor of \textcolor{black}{a} quasi-BIC 
on the disk diameter. Right: measured reflectance spectra of nano\textcolor{black}{disks with different} 
diameter\textcolor{black}{s}. 
Adapted from Melik-Gaykazyan et al~\cite{melik2021fano}.}  \label{fig:3d1}
\end{figure}

For individual subwavelength resonators, 
genuine nonradiative states require extreme \textcolor{black}{values of permittivity,} 
\textcolor{black}{tending} toward infinity or zero~\cite{monticone2014embedded, silveirinha2014trapping, hayran2021capturing} or \textcolor{black}{imitating periodic} 
boundary conditions with metallic waveguides~\cite{lepetit2010resonance, lepetit2014controlling, jacobsen2021boundary}. In realistic individual resonators, there \textcolor{black}{is} 
always an infinite number of radiation channels, which limits the Q factor substantially. However, the concept of quasi-BICs allows 
reaching \textcolor{black}{almost} nonradiative states for individual dielectric resonators. Using the parameter-tuning approach originally developed by Friedrich and Wintgen~\cite{friedrich1985physical} and later used for some extended geometries~\cite{wiersig2006formation}, one can \textcolor{black}{create} 
high-Q quasi-BICs in compact geometries at subwavelength scales.  In \textcolor{black}{a} very recent work~\cite{rybin2017optical}, 
\textcolor{black}{implementation of} quasi-BICs, \textcolor{black}{called there} supercavity modes, \textcolor{black}{was proposed} in individual dielectric resonators via continuous tuning of the resonator’s aspect ratio. Such parameter tuning enables destructive interference and strong coupling of 
\textcolor{black}{two} leaky modes (radial and axial) 
when their frequencies come close. 
Importantly, the modes forming a quasi-BIC belong to the same resonator \textcolor{black}{should have the same symmetry}.

The concept of \textcolor{black}{the mentioned} supercavity mode 
is shown in Fig.~\ref{fig:3d1}(a). The map demonstrates the \textcolor{black}{contribution of the modes with} 
zero azimuthal index \textcolor{black}{to the scattering cross section of} 
a high-index dielectric disk with respect to dimensionless frequency and disk aspect ratio. The radial and axial mode\textcolor{black}{s} interact strongly in the vicinity of the avoided resonance crossing, which results in sharp narrowing of the linewidth for one of the modes - supercavity mode. The second mode demonstrate\textcolor{black}{s a} 
broadening of the linewidth according to \textcolor{black}{the} general properties of open non-Hermitian systems~\cite{cao2015dielectric} [see also Fig.~\ref{fig:2a1}(b)]. The proposed hybrid mode profiles \textcolor{black}{are}  
a combination of radial and axial oscillations in the vertical cross-section and a uniform azimuthally symmetric distribution in the horizontal cross-section. The formation of quasi-BICs 
\textcolor{black}{can be identified by the special features of} the lineshape in scattering \textcolor{black}{spectrum}. For a high-index disk with \textcolor{black}{a} permittivity of $80$, the lineshape changes from \textcolor{black}{an} asymmetric Fano profile to \textcolor{black}{a} symmetric Lorentzian in the vicinity of supercavity mode formation, as shown in Fig.~\ref{fig:3d1}(b). This \textcolor{black}{change} 
corresponds to diverging Fano asymmetry parameter $q$. Such behaviour was later explained within the framework of 
interference of \textcolor{black}{different} modes with very similar far-field profile\textcolor{black}{s}~\cite{bogdanov2019bound,chen2019multipolar}. In the case in Fig.~\ref{fig:3d1}(a), both interacting modes are dominated by magnetic dipolar contribution. \textcolor{black}{The dipole contributions can cancel each other} 
via mode interference, which makes the next allowed \textcolor{black}{multipole (in this case,} magnetic octupol{e})
contribution dominant. The formation of quasi-BICs with nonzero azimuthal indices is also possible, but \textcolor{black}{it} leads to lower Q factors due to lower symmetry of the field profile.

The first experimental observation of \textcolor{black}{quasi-BICs} 
was carried out \textcolor{black}{quite} 
recently in microwave~\cite{odit2021observation} and near-IR~\cite{koshelev2020dielectric, melik2021fano} \textcolor{black}{ranges}. Figure~\ref{fig:3d1}(b) shows the measured spectral map of the reflection coefficient $1 - |S_{11}|$ 
\textcolor{black}{for} a single ceramic disk with permittivity of $44.8$ from Ref.~\cite{odit2021observation}. The excitation was performed \textcolor{black}{in the near field} with a loop antenna, \textcolor{black}{which selectively excites the modes with zero azimuthal index.} 
The spectra reveal four avoided resonance crossings corresponding to formation of quasi-BICs. \textcolor{black}{The radiative Q factor of quasi-BICs reaches $200000$, but due to the absorption in the ceramic material,} the maximal measured unloaded total Q factor \textcolor{black}{is} 
$12500$. 
The observation of supercavity modes in near-IR was carried out for isolated AlGaAs nanodisks on engineered three-layer substrate with a reflective layer of ITO in the middle~\cite{koshelev2020dielectric, melik2021fano}. The SEM image of individual nanodisks is shown in \textcolor{black}{the} upper panel of Fig.~\ref{fig:3d1}(c). The 
measured spectra and extracted Q factor for a range of the disk diameters is shown in the lower panel of Fig.~\ref{fig:3d1}(c). The maxim\textcolor{black}{um} 
measured Q factor is \textcolor{black}{approximately} 
$180$ for a disk with $930$~nm diameter. The spectra reveal that the lineshape is Fano-like and changes \textcolor{black}{its} asymmetry in the vicinity of the quasi-BIC, 
as was predicted theoretically for a \textcolor{black}{single} 
disk \textcolor{black}{in vacuum}. To match the mode symmetry \textcolor{black}{and excitation,} a tightly focused azimuthally polarized Gaussian beam was used. 

Later, several generalizations were suggested for quasi-BICs in textcolor{black}{non-periodic finite resonators.} 
\textcolor{black}{For example, } two coaxial disks were suggested as a structure supporting quasi-BICs with much higher values of Q factor \textcolor{black}{than that of a single disk}~\cite{pichugin2019interaction}. Additionally, Bragg reflectors were suggested to further increase the Q factor of an individual quasi-BIC disk~\cite{kolodny2020q}. The multipolar classification  of eigenmodes in  
\textcolor{black}{single} resonators of different shapes was suggested later, based on the symmetry group theory~\cite{gladyshev2020symmetry}. These results lead to prediction of quasi-BICs in dielectric spheroids~\cite{bulgakov2021exceptional}, triangular prisms~\cite{gladyshev2020symmetry}, bar-shaped microcavities~\cite{huang2021pushing} and resonators of \textcolor{black}{arbitrary} 
shapes~\cite{yan2020shape}.

\subsection{Asymmetric metasurfaces \textcolor{black}{and quasi-BICs}}

Recently, it was demonstrated that 
broken-symmetry dielectric metasurfaces demonstrate \textcolor{black}{sharp peaks} 
in the normal incidence 
transmission \textcolor{black}{spectra, which are} associated with 
quasi-BICs \textcolor{black}{excitation}~\cite{koshelev2018asymmetric}. It was shown that the in-plane asymmetry induces an imbalance of the interference between \textcolor{black}{counter}
-propagating leaky waves, \textcolor{black}{resulting in the formation of} 
a quasi-BIC.
This effect was observed in a variety of metasurfaces in different frequency ranges and was connected to \textcolor{black}{electromagnetic field}
-induced transparency~\cite{singh2011observing}, tunable high-Q 
resonances~\cite{zhang2014fano}, trapped-mode resonances~\cite{fedotov2007sharp}, broken-symmetry Fano metasurfaces~\cite{campione2016broken,vabishchevich2018enhanced} and dark modes~\cite{jain2015electric}. 

\begin{figure}[t]
\includegraphics[width=0.9\linewidth]{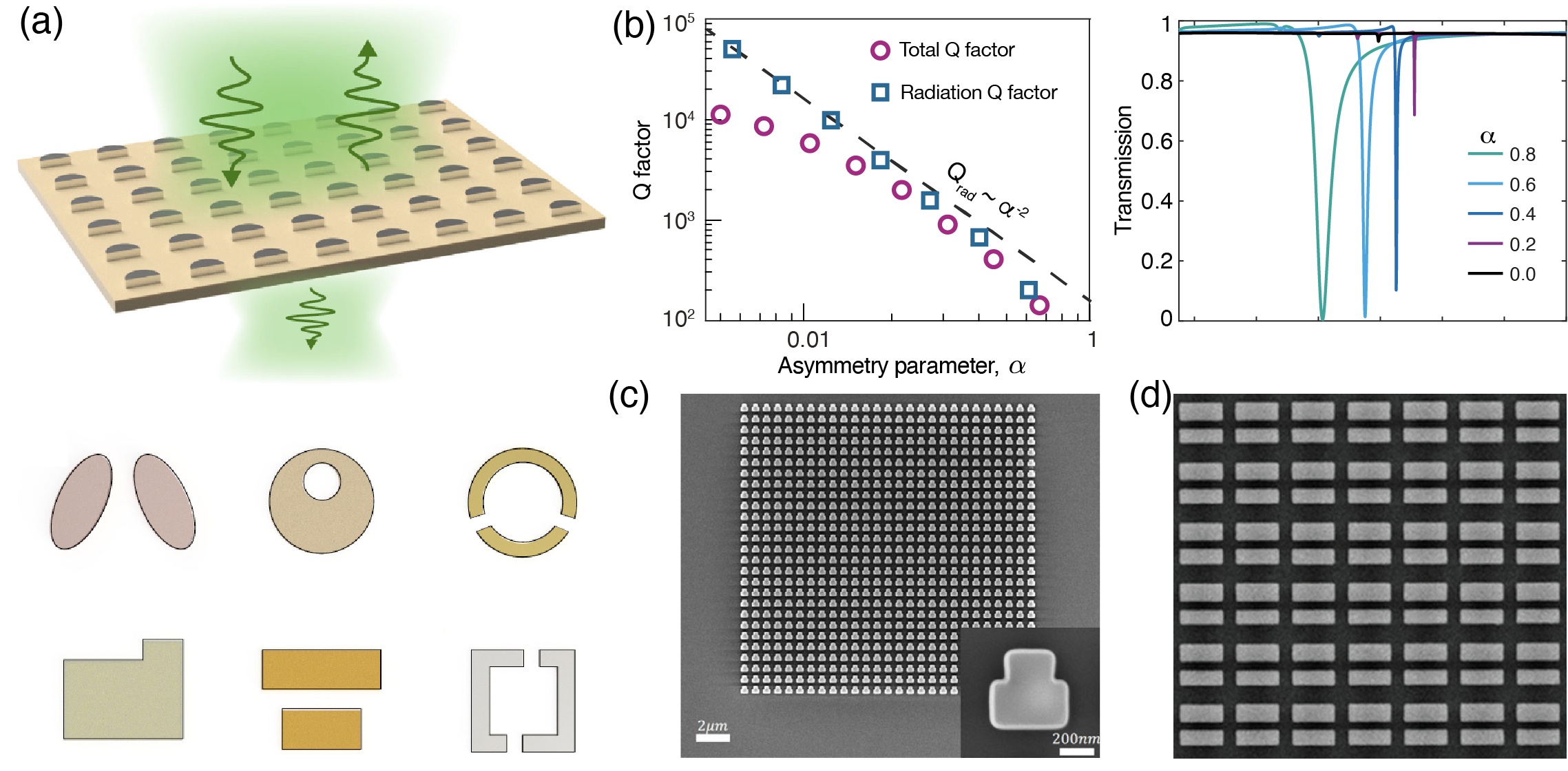}
\caption{(a) Schematic for the scattering of light by a metasurface, and designs of the unit cells of asymmetric metasurfaces with a broken in-plane inversion symmetry of constituting meta-atoms supporting sharp resonances. (b) Typical dependence of the total (circles) and radiative (squares) Q factor of the quasi-BIC on the meta-atom asymmetry parameter $\alpha$ and characteristic evolution of the transmission spectra by changing $\alpha$. (c,d) SEM images of silicon metasurfaces used in recent experimental demonstrations of quasi-BICs with a very high Q factor of \textcolor{black}{approximately} 
18500 (c) and 750 (d). Adapted from Koshelev et al.~\cite{koshelev2018asymmetric,koshelev2020dielectric}}\label{fig:4a1}
\end{figure}

Figure~\ref{fig:4a1}(a) shows the scattering of light by an asymmetric metasurface. The radiative Q factor of quasi-BICs in asymmetric metasurfaces follows the typical inverse quadratic dependence on the meta-atom asymmetry parameter $\alpha$,~\cite{koshelev2018asymmetric} as shown in Fig.~\ref{fig:4a1}(b), left panel. In the regime of transmission (or reflection), the quasi-BICs manifest themselves as sharp asymmetric Fano resonances \textcolor{black}{whose} 
width and depth decreases with the decrease of $\alpha$, as shown in Fig.~\ref{fig:4a1}(b), right panel. The total Q factor of the quasi-BIC mode is limited by other losses (see Section \textcolor{black}{'Losses and Q factor of quasi-BICs'}). Due to parasitic effect of other losses, the maximal field enhancement can be achieved not for minimal radiative losses, but in the regime of optimal (critical) coupling when the rates of radiative and parasitic losses are equal, \textcolor{black}{$Q_\text{abs}=Q_\text{rad}$}~\cite{seok2011radiation, koshelev2019nonlinear}. Very recently, quasi-BICs with giant values of the Q factor in Si metasurfaces were experimentally demonstrated for different designs of meta-atoms. Figures~\ref{fig:4a1}(c) and \ref{fig:4a1}(d) show the SEM images of two metasurfaces hosting the quasi-BICs with a Q factor of about 18500,~\cite{liu2019high} and 750,~\cite{ndao2020differentiating} achieved by smart engineering of radiative losses and advanced electron-beam lithography techniques. Later, it was shown that even a true BIC can exist in an asymmetric metasurface in specific conditions~\cite{han2021extended}. Metasurfaces with strong asymmetry were used to create near-unity chiral response~\cite{gorkunov2020metasurfaces,gorkunov2021bound,overvig2021chiral}.

\section{\textcolor{black}{BIC applications: optical detection, laser generation, nonlinear optics}}

\begin{figure}[t]
\includegraphics[width=0.9\linewidth]{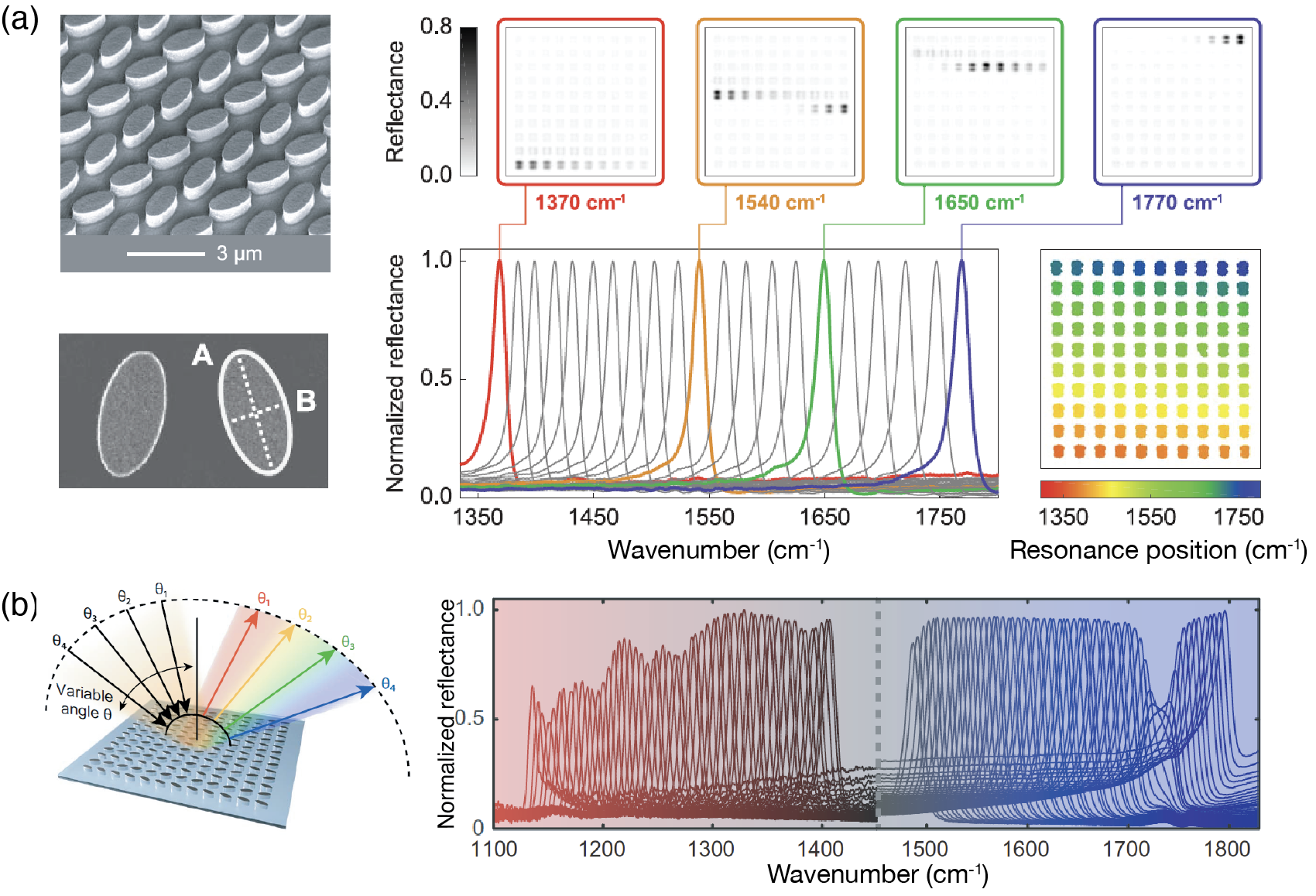}
\caption{{\textcolor{black}{(a) SEM image of a metasurface and its meta-atom. Reflection images of a pixel metasurface obtained for four given wavelengths in the mid-IR range. Normalized reflection spectra for 21 out of 100 metapixels. Based on A.~Tittl et al.\cite{tittl2018imaging}. (b) A germanium-based dielectric metasurface supporting quasi-BIC with angle multiplexing. Reflection spectra after deposition of a thin film of polymethyl methacrylate with centrifugation. Based on the work of A.~Leitis and colleagues\cite{leitis2019angle}}}.}
\label{fig:4b1}
\end{figure}

\textcolor{black}{BICs and quasi-BICs are widely used for various applications in photonic crystal slabs, waveguides, single subwavelength particles and other platforms. BIC-based applications include filtering~\cite{foley2014symmetry,foley2015normal,cui2016normal,doskolovich2019integrated}, lasing~\cite{gentry2014dark,kodigala2017lasing, midya2018coherent,ha2018directional,wu2020room,azzam2021single,huang2020ultrafast,muhammad2021optical,hwang2021ultralow,dyakov2021photonic,yang2021low}, magnetophotonics~\cite{ignatyeva2020bound,chernyak2020bound,Zakharov2020,chen2019strong}, detection of biological objects~\cite{zhen2013enabling,sun2016high,wang2018optofluidic,meudt2020hybrid,romano2018surface,romano2019tuning,ndao2020differentiating,wang2021ultrasensitive, tittl2018imaging,leitis2019angle, yesilkoy2019ultrasensitive,jahani2021imaging}, nonlinear generation and self-action~\cite{bulgakov2008bound,bulgakov2010bound,ndangali2013resonant,bulgakov2014robust,bulgakov2015all,pichugin2015frequency,pichugin2016self,wang2017improved,yuan2017strong,krasikov2018nonlinear,yuan2020excitation,deka2018microscopic,chukhrov2021excitation,maksimov2020optical,koshelev2019meta,koshelev2019nonlinear,zograf2020high,sinev2021observation}, vortex generation~\cite{bahari2017integrated,doeleman2018experimental,wang2020generating}, on-chip photonic devices~\cite{bykov2019bound,yu2019photonic,doskolovich2019integrated,yu2020high,wang2020bound}, switches driven by external voltage~\cite{henkel2021electrically},  active THz devices~\cite{han2019all}, optical tuning of halcogenide  metasurfaces~\cite{mikheeva2019photosensitive}, enhancement of chiral nonlinear response~\cite{gandolfi2021near}, polaritonics~\cite{PhysRevB.98.161113,kravtsov2019control,cao2020normal,qin2021strong,Zheng2021,Zong2021,AlAni2021}, and harmonic generation in hybrid structures with monolayers of transition metal dichalcogenides~\cite{bernhardt2020quasi, lochner2020hybrid}.
Moreover, BICs in single nanoparticles were used to achieve a record-high efficiency of second-harmonic and higher harmonic generation~\cite{carletti2018giant,carletti2019high,koshelev2020dielectric,kolodny2021enhancement}, generation of quantum-entangled photons~\cite{poddubny2018nonlinear}, and low-threshold lasing~\cite{mylnikov2020lasing}. }

\textcolor{black}{We will consider several applications in more detail. Passive photonic structures supporting BICs recently were used to increase the efficiency of biological objects' detection. In the work~\cite{tittl2018imaging}, Tittl et al. implemented a nanophotonic sensor of biomolecules in the mid-IR range, based on the reflection from a dielectric matrix of metasurfaces supporting BICs. In Fig.~\ref{fig:4b1}(a), the left panel shows SEM images of an asymmetric silicon metasurface and its meta-atoms. The proposed structure supports quasi-BICs in the mid-IR range with a Q factor of approximately 200. The right panel shows reflection images of a pixel metasurface obtained for four given wavelengths and normalized reflection spectra for 21 out of 100 metapixels. Due to the narrow quasi-BIC resonance in the reflection spectrum and its tunability with changes in the geometric dimensions of the metasurface, a method based on this structure for distinguishing the absorption spectra of various molecules was demonstrated. In the work~\cite{leitis2019angle} by Leitis et al., an asymmetric silicon metasurface with BICs was used for angle multiplexing of spectra, which also allowed a convenient distinguishing of absorption spectra of various biomolecules. Figure~\ref{fig:4b1}(b) shows a metasurface multiplexing scheme and reflection spectra from the structure. In another series of works by Romano et al.~\cite{romano2018label, romano2018optical,romano2018surface}, the sensitivity of sensors based on photonic crystal waveguides with BICs was studied to changes in the refractive index of the external medium, and high sensitivity was experimentally shown to be achieved due to BICs. In a recent work by Jahani et al~\cite{jahani2021imaging}, BIC-based silicon metasurfaces with a complex unit cell were developed for the detection of extracellular vesicles of cancer cells.}

\begin{figure}[t]
\includegraphics[width=0.97\linewidth]{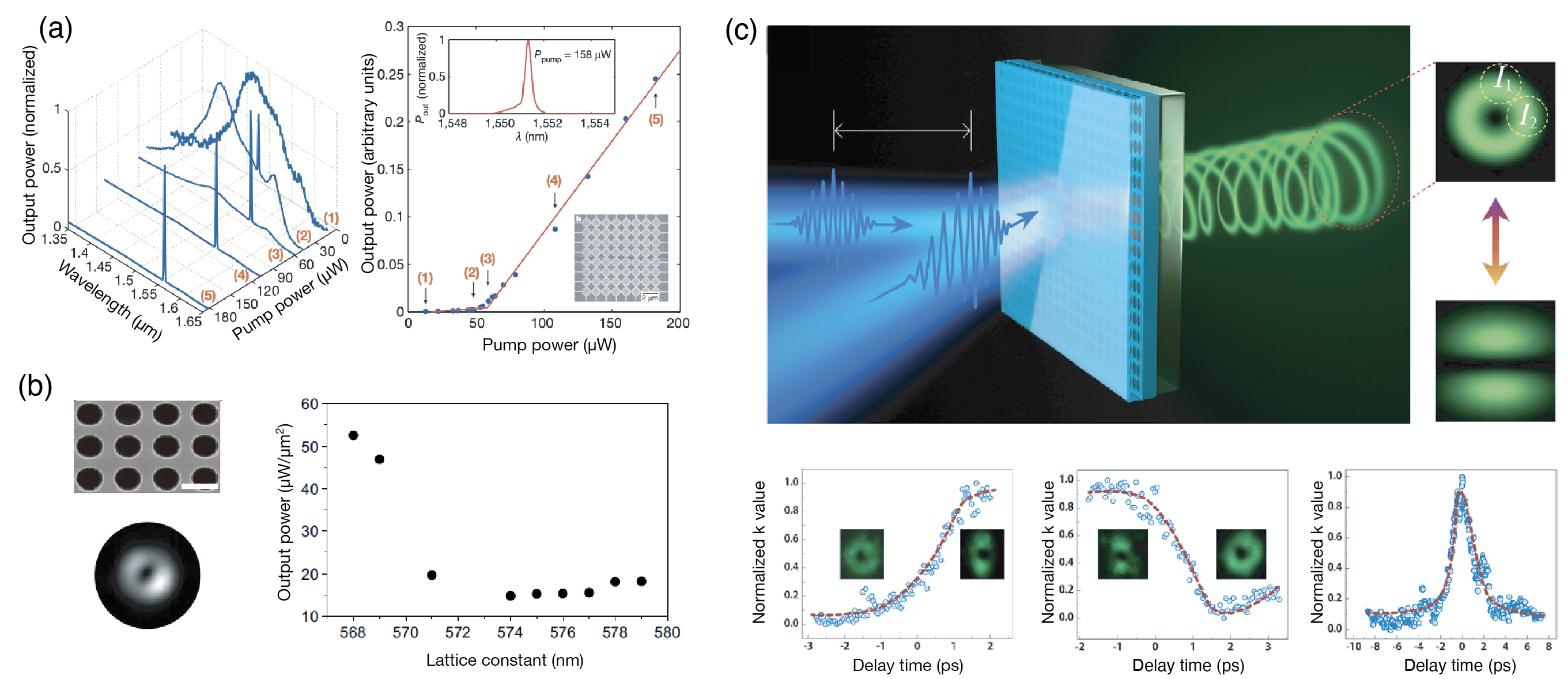}
\caption{{\textcolor{black}{(a) Change in the normalized output power as a function of wavelength and pump power for a metasurface. Output power as a function of the average pump power (light curve) at the lasing wavelength. An image of the $16 \times16$ metasurface is shown in the inset. Based on Kodigala et al.\cite{kodigala2017lasing}. (b) SEM image of a photonic crystal waveguide and measured far-field profile for a super-BIC laser. The value of the lasing threshold normalized to the pumping area (5.4 $\mu$m in size) as a function of the lattice parameter. Based on M.~Hwang et al.~\cite{hwang2021ultralow}. (c) Top: Schematic of a two-beam pumping experiment. The insets show far-field radiation patterns of a perovskite photonic structure for symmetric and asymmetric pump beam profiles. Bottom: Evolution from vortex laser generation to linearly polarized generation and the reverse process. Transition from an annular beam to a two-lobe beam and back on the scale of several picoseconds. Based on the work of Huang et al.~\cite{huang2020ultrafast}.}}}\label{fig:4b2}
\end{figure}

\textcolor{black}{Photonic structures with BICs are widely used in active photonics, including for laser generation. Figure~\ref{fig:4b2}(a) shows the 
output power spectra depending on the wavelength and pump power for a $16 \times16$ metasurface supporting BICs in the near-IR range from~\cite{kodigala2017lasing}. In this work, the Authors used a tunable accidental BIC with a wavelength of $1550$~nm. The right panel of Fig.~\ref{fig:4b2}(a) shows the dependence of the output power on the pump power at resonance. As the pump power is increased to $60$ $\mu$W, a distinct peak is observed in the emitted power spectra at the BIC wavelength. The inset shows a SEM image of the structure. In a recent paper~\cite{hwang2021ultralow}, lasing from the so-called super-BIC, i.e. several BICs combined in the momentum space and having the same frequency, was studied. A distinctive feature of a super-BIC is a higher stability of the mode against deviations from periodicity and imperfections in the surface of the structure. The left panel of Fig.~\ref{fig:4b2}(b) shows a SEM image of an InGaAsP photonic crystal waveguide and the measured far-field profile for a mode of the super-BIC laser. Laser generation in the structure is achieved through the use of quantum dots in the layer. As shown in Fig.~\ref{fig:4b2}(b), the spot in the Fourier space has a small angular divergence, which is a hallmark of strongly confined modes.
The right panel of Fig.~\ref{fig:4b2}(b) shows the lasing threshold normalized to the pumping area ($5.4~\mu$m in size) as a function of the lattice parameter of the structure. The super-BIC state is realized at a lattice parameter of about $573$~nm: in this regime, the lasing threshold is the lowest and has a record low value among all BIC lasers and lasers based on topologically protected states~\cite{hwang2021ultralow}. In the work of Huang et al.~\cite{huang2020ultrafast}, vortex microlasers were studied, based on perovskite photonic crystal waveguides supporting BICs in the visible range, for ultrafast optical switching at room temperature. The Authors have experimentally demonstrated switching between vortex beam generation and linearly polarized beam generation, with a characteristic switching time from 1 to 1.5 picoseconds and a record low power consumption. The top panel of Fig.~\ref{fig:4b2}(c) shows the scheme of the experiment on two-beam pumping of a perovskite photonic structure. The insets show far-field radiation patterns for symmetric and asymmetric pump beam profiles. The lower panel shows the transition from vortex laser generation to linearly polarized generation and the reverse process on the scale of several picoseconds. S.~Ha et al.~\cite{ha2018directional} studied a laser based on a gallium arsenide metasurface supporting a BIC at a wavelength of about $825$~nm.
Azzam et al.~\cite{azzam2021single} studied single-mode and multimode lasing in the visible range based on a similar titanium dioxide metasurface coated with a thin layer of an organic dye. In this work, it was experimentally shown that directed radiation can be achieved by adjusting the period of the metasurface due to the BIC properties. In a recent work by Yang et al.~\cite{yang2021low}, lasing based on an asymmetric Si$_3$N$_4$ metasurface coated with a rhodamine 6G dye with strong fluorescent properties was studied. In this work, lasing at a wavelength of approxmately 600~nm was experimentally demonstrated due to the excitation of a quasi-BIC. S.Dyakov et al.~\cite{dyakov2021photonic} showed a significant increase in the photoluminescence from a silicon photonic-crystal waveguide with germanium nanoislands, which supports BICs in the near-IR range.}

\begin{figure}[t]
\includegraphics[width=0.85\linewidth]{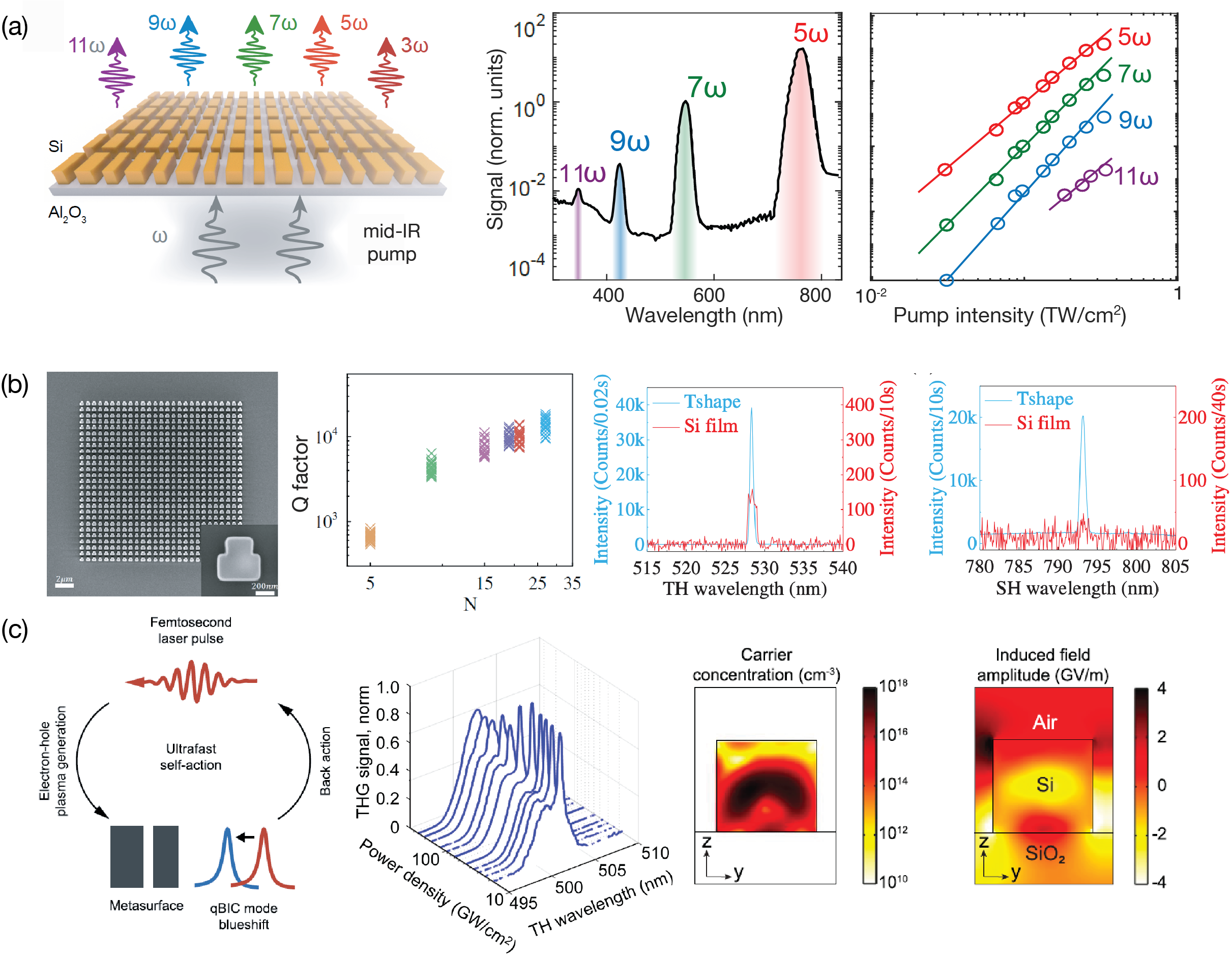}
\caption{{\textcolor{black}{(a) The spectrum of the signal from the 3rd to 11th harmonics from an asymmetric silicon metasurface with BICs, and the dependence of the harmonics' power on the pump pulse power. Based on G. Zograf et al.\cite{zograf2020high}. (b) Statistics of the measured Q factors of the BICs for all the fabricated silicon metasurfaces with different sizes. The spectra of the second and third harmonics for a metasurface and an unstructured film. Based on J. Liu et al.\cite{liu2019high}. (c) Scheme of ultrafast self-action of a pulse through a BIC. The spectrum of the third harmonic depending on the pump power. Carrier concentration and induced field amplitude inside the nanostructure at a pump intensity of 240 GW/cm$^2$. Based on the work of I. Sinev and colleagues\cite{sinev2021observation}.}}} \label{fig:4b3}
\end{figure}

\textcolor{black}{BICs and quasi-BICs are studied most actively in nonlinear optics and photonics applications, mainly in order to enhance nonlinear generation and to observe the self-action of an exciting pulse. In a series of works by K. Koshelev et al.~\cite{koshelev2019meta,koshelev2019nonlinear}, optical second- and third-harmonic generation was studied in nonlinear dielectric metasurfaces with asymmetric unit cell, supporting quasi-BICs with Q factor depending on the asymmetry. In particular, in the work~\cite{koshelev2019nonlinear}, the fabricated metasurface had a low Q factor of approximately 160 due to the presence of strong surface roughness. In this work, it was shown theoretically and experimentally that for such non-perfect photonic structures, the highest efficiency of harmonic generation is achieved not at the highest Q factor, but in the critical coupling regime, when the radiative Q factor is equal to the Q factor related to all the other types of losses.
It was also shown in the work that the critical coupling regime can be achieved by changing the asymmetry parameter of the meta-atom, which should be taken into account in the design of resonant nonlinear metasurfaces. G. Zograf et al.~\cite{zograf2020high} studied the generation of higher-order odd harmonics (3,5,7,9,11) from an asymmetric silicon metasurface, shown schematically in the left panel of Fig.~\ref{fig:4b3}(a). The right panel of Fig.~\ref{fig:4b3}(a) shows the spectra of optical harmonics in the regime of excitation with $100$~fs pulses and the dependence of the output power on the pump power for an optimized metasurface, using the critical coupling criterion.
The dependence of the output power from pump powers in the range of $0.03-0.3$~TW/cm$^2$ is defined by the same law, regardless of the harmonic number, which allows a conclusion that the structure operates in a nonperturbative regime. In this paper, the transition to the nonperturbative regime is explained through the mechanism of free-carrier generation in silicon due to multiphoton absorption. J.Liu et al.~\cite{liu2019high} studied asymmetric silicon metasurfaces with T-shaped meta-atoms, shown in Fig.~\ref{fig:4b3}(b). A record-high Q factor of a BIC of over 20000 was obtained in this work for a structure with 26$\times$26 periods, achieved due to the high quality of nanolithography and special features of the structure design. In this work, the spectra of the optical second- and third-harmonic signals were measured, shown in the right panel of Fig.~\ref{fig:4b3}(b), and at the BIC wavelength, the harmonic signal increases by several times. I. Sinev et al.~\cite{sinev2021observation} studied an asymmetric silicon metasurface for third-harmonic generation at high pump power intensities of the order of $0.3$~TW/cm$^2$. It was demonstrated in the work that at a pump intensity higher than $240$~GW/cm$^2$, self-action of a pulse occurs due to the generation of free charge carriers in the multiphoton absorption mechanism, schematically shown in the left panel of Fig.~\ref{fig:4b3}(c). The calculated induced field amplitude and charge carrier concentration at a peak intensity of $240$~GW/cm$^2$ reached $3$~GW/m and $10^{18}$~cm$^{-3}$, respectively, as shown in the right panel of Fig.~\ref{fig:4b3}(c).}


\subsection{Conclusion}

In conclusion, bound states in the continuum have a long history in optics and radiophysics. 
\textcolor{black}{Today, this area reaches a new level} due to the fast development of the physics of metasurfaces, 2D materials, nonlinear nanophotonics, flat optics, and related directions. 
BICs can be observed in a wide variety of photonic structures, including metasurfaces, photonic crystal slabs, high-contrast gratings, corrugated planar waveguides and fibers, ridge waveguides, linear chains, and many others.
\textcolor{black}{A relatively recent discovery is a new class of BICs in Bragg resonators with an anisotropic defect layer, proposed and experimentally studied in the works~\cite{timofeev2018optical,pankin2020fano,pankin2020one,wu2021quasi}.}
Today, the physics of BICs is developing in acoustics~\cite{tong2020observation,deriy2022bound,lyapina2015bound,huang2020extreme}. 
BIC\textcolor{black}{s are} 
an illustrative example of how an idea suggested in one \textcolor{black}{area}  
of physics a century ago today affects many other fields and \textcolor{black}{is already used in} 
multiple practical applications.      

\begin{acknowledgement}
The authors acknowledge financial support from the Russian Foundation for Basic Research (grant no. 20-12-50314). The Authors thank A. Sadreev, E. Bulgakov, D. Maksimov, E. Bezus, D. Bykov for fruitful discussions.  
\end{acknowledgement}

\bibliography{Literature-final}

\end{document}